\documentclass[fleqn,usenatbib]{mnras}

\usepackage{newtxtext,newtxmath}
\usepackage{hyperref}

\usepackage[T1]{fontenc}
\usepackage{longtable}
\usepackage{caption}

\DeclareRobustCommand{\VAN}[3]{#2}
\let\VANthebibliography\thebibliography
\def\thebibliography{\DeclareRobustCommand{\VAN}[3]{##3}\VANthebibliography}

\usepackage{graphicx}	
\usepackage{amsmath}	

\title[LMC's Perturbations in the Milky Way Halo]{LMC-induced Perturbations in the Milky Way Halo:\\ I. HaloDance Simulation Suite and Observational Forecasts}

\author[Sheng et al.]{
Yanjun Sheng,$^{1}$\thanks{E-mail: Yanjun.Sheng@anu.edu.au}
Yuan-Sen Ting,$^{2,3}$
Xiang-Xiang Xue$^{4,5}$
\\
$^{1}$Research School of Astronomy $\&$ Astrophysics, Australian National University, Cotter Rd., Weston, ACT 2611, Australia\\
$^{2}$Department of Astronomy, The Ohio State University, Columbus, OH 43210, USA\\
$^{3}$Center for Cosmology and AstroParticle Physics (CCAPP), The Ohio State University, Columbus, OH 43210, USA\\
$^{4}$National Astronomical Observatories, Chinese Academy of Sciences, Beijing 100101, People’s Republic of China\\
$^{5}$Institute for Frontiers in Astronomy and Astrophysics, Beijing Normal University, Beijing 102206, Peoples Republic of China
}

\date{Accepted XXX. Received YYY; in original form ZZZ}

\pubyear{\the\year{}}

\begin{document}
\label{firstpage}
\pagerange{\pageref{firstpage}--\pageref{lastpage}}
\maketitle

\begin{abstract}
The gravitational interaction between the Milky Way (MW) and Large Magellanic Cloud (LMC) perturbs the MW halo's density and kinematics, encoding information about both galaxies’ masses and structures. We present a suite of 2,848 high-resolution ($10^{7}$ particles) N-body simulations that systematically vary the mass and shape of both galaxies’ haloes. We model how the mean velocities and velocity dispersions of halo stars (30–120 kpc) depend on system parameters, and forecast constraints achievable with current and future observations. With fiducial values of $M_{\mathrm{MW}} = 0.7\times10^{12}M_\odot$, $M_{\mathrm{LMC}} = 1.5\times10^{11}M_\odot$, $c=9.415$ (MW halo concentration), $q=1.0$ (halo flattening), and assuming Gaia DR3-level astrometry, 20 km/s radial velocity precision, 10\% distance precision, and a sample of 4,000 RR Lyrae stars, we achieve $1\sigma$ uncertainties of $0.11\times10^{12}M_\odot$ in $M_{\mathrm{MW}}$, $2.33\times10^{10}M_\odot$ in $M_{\mathrm{LMC}}$, 2.38 in $c$, and 0.06 in $q$. These correspond to fractional uncertainties of 11\%, 16\%, 25\%, and 6\%, respectively, relative to fiducial values. Improved Gaia proper motions (DR5) yield modest gains (up to 14\%), while adding radial velocities improves constraints by up to 60\% relative to using Gaia astrometry alone. Doubling the sample size to $\sim$8,000 stars provides additional 30\% improvements, whereas reducing distance uncertainties has minimal impact ($\lesssim$10\%). Mean velocities trace LMC-induced perturbations, while velocity dispersions constrain the MW halo properties, jointly breaking degeneracies. Our results demonstrate that combining Gaia astrometry with large spectroscopic surveys will enable precise characterization of the MW-LMC system. This methodology paper establishes the framework for interpreting observations; future work will apply these tools to existing spectroscopic datasets. The full simulation suite, \textit{HaloDance}, will be made publicly available at \href{https://github.com/Yanjun-Sheng/HaloDance}{\texttt{github.com/Yanjun-Sheng/HaloDance}}.
\end{abstract}


\section{Introduction}

The Large Magellanic Cloud (LMC) is the most massive satellite galaxy of the Milky Way (MW), with a virial mass estimated at $0.8$–$2.5 \times 10^{11} {\rm M}_{\odot}$, representing 10-25\% of the MW's mass \citep{2016ARA&A..54..529B,2020SCPMA..6309801W,2021ApJ...923..149S,2024A&A...688A.147F}. The mass of the LMC has been constrained through multiple independent methods: the LMC's rotation curve \citep{2014ApJ...781..121V}, kinematics of LMC globular clusters \citep{2024ApJ...963...84W}, shapes and kinematics of stellar streams in the MW \citep{2019MNRAS.487.2685E,2021ApJ...923..149S,2021MNRAS.501.2279V,2022MNRAS.511.2610C,2023MNRAS.521.4936K}, and the timing argument with M31 and the nearby Hubble flow \citep{2016MNRAS.456L..54P}.

The gravitational interaction between such a massive satellite and its host galaxy produces distinct dynamical signatures in the MW's dark matter and stellar halos. These signatures manifest on both small and large scales, with amplitudes that scale with the masses of both galaxies, providing a powerful tool for constraining the properties of the MW-LMC system \citep{2015ApJ...802..128G,2016MNRAS.457.2164O,2018MNRAS.473.1218L,2018MNRAS.481..286L,2019ApJ...884...51G,2020MNRAS.494L..11P,2021MNRAS.506.2677E,2023Galax..11...59V,2024OJAp....7E..50K}.

On small scales, the LMC generates a dynamical friction wake—an overdensity of dark matter and stars trailing its past trajectory \citep{1943ApJ....97..255C,2009MNRAS.400.1247C,2016MNRAS.457.2164O,2019ApJ...884...51G,2021ApJ...916...55T,2023ApJ...954..163F}. The amplitude of this wake depends on the total mass, shape, and density profile of the Galactic halo \citep{2016ARA&A..54..363D,2017MNRAS.464.3825P,2019ApJ...884...51G,2024MNRAS.534.2694S}, increasing with higher velocity anisotropy or larger LMC infall mass \citep{2019ApJ...884...51G,2021ApJ...919..109G,2023ApJ...954..163F,2024A&A...688A.147F}. This wake has been observationally detected using K giants and Blue Horizontal Branch (BHB) stars \citep{2021Natur.592..534C,2024A&A...690A.166A} and linked to the previously known Pisces plume \citep{2019MNRAS.488L..47B}.

On larger scales, the LMC induces a global reflex motion in the MW. The Galactic disk and inner halo, with their shorter dynamical timescales, respond faster to the LMC's gravitational pull than the outer halo. This differential response creates a bulk motion of the inner MW relative to its outskirts. Simulations predict that the inner MW accelerates toward the LMC while the outer halo remains nearly stationary, producing characteristic kinematic signatures: a north-south dipole in radial velocities and an all-sky positive bias in latitudinal velocities \citep{2018MNRAS.473.1218L,2019ApJ...884...51G,2020MNRAS.494L..11P,2020ApJ...898....4C,2021ApJ...919..109G,2023Galax..11...59V,2024MNRAS.527..437V,2024MNRAS.534.2694S}.

These predicted signatures have been detected in various stellar populations. The dipole pattern has been observed in K-giants and BHB stars beyond 50 kpc \citep{2021MNRAS.506.2677E,2024arXiv240601676C}, though \citet{2024arXiv241009149B} identified an additional monopole component suggesting more complex kinematics. Multiple studies have measured the reflex motion's amplitude and direction \citep{2021NatAs...5..251P,2024MNRAS.531.3524Y,2024arXiv240601676C,2024arXiv241009149B}, though discrepancies exist between observational results and N-body predictions regarding the apex direction \citep{2019ApJ...884...51G,2024MNRAS.527..437V}.

The relationship between these dynamical effects and galaxy masses has enabled direct mass measurements through kinematic modeling. By comparing observed velocity fields with simulations, studies have constrained the LMC mass to values ranging from $1.5 \times 10^{11} \mathrm{M}_{\odot}$ \citep{2021MNRAS.506.2677E} to $2.1 \times 10^{11} \mathrm{M}_{\odot}$ \citep{2024MNRAS.534.2694S}, with some analyses requiring masses above $1.8 \times 10^{11} \mathrm{M}_{\odot}$ \citep{2024arXiv240601676C}. Stellar streams provide complementary constraints: the Orphan-Chenab stream yields LMC masses of $1.3$-$1.9 \times 10^{11} \mathrm{M}_{\odot}$ \citep{2019MNRAS.487.2685E,2021ApJ...923..149S,2023MNRAS.521.4936K}, while the Sagittarius stream suggests $\sim 1.3 \times 10^{11} \mathrm{M}_{\odot}$ \citep{2021MNRAS.501.2279V}.

The orbital history of the LMC remains debated. Most studies favor a first-infall scenario with a recent pericenter passage at approximately 50 kpc \citep[e.g.,][]{2007ApJ...668..949B,2010ApJ...721L..97B,2013ApJ...764..161K,2020ApJ...893..121P,2024MNRAS.534.2694S}, supported by evidence from star formation history \citep{2009AJ....138.1243H,2014MNRAS.438.1067M,2022MNRAS.513L..40M}, stellar streams \citep{2020ApJ...905L...3Z,2022MNRAS.514.1266P,2023ApJ...956..110C}, and dynamical effects \citep{2021Natur.592..534C,2023ApJ...954..163F,2024MNRAS.534.2694S}. However, a second passage scenario remains possible \citep{2024MNRAS.527..437V}.

Despite the wealth of kinematic information available, most studies have relied on limited sets of fiducial simulations, restricting analyses to qualitative features or individual tracers like stellar streams. This limitation has prevented robust posterior inference for MW-LMC parameters across the full parameter space. Our previous work \citep{2024MNRAS.534.2694S} demonstrated that the LMC's dynamical effects depend sensitively on both galaxies' properties, supporting a massive first-infalling LMC even when considering parameter variations. However, a comprehensive exploration of the parameter space with sufficient resolution to enable quantitative inference has been lacking.

In this study, we present a comprehensive suite of 2,848 high-resolution ($10^7$ particles) simulations that systematically explore the MW-LMC parameter space, focusing exclusively on models in which the LMC follows a first-infall orbit. We model the mean velocity and velocity dispersion of halo stars within 30–120 kpc as functions of four key parameters: MW mass, LMC mass, MW halo concentration, and halo shape. We then evaluate how observational uncertainties—from current Gaia DR3 precision to expected DR5 improvements—and sampling noise affect parameter constraints. This methodology paper establishes the theoretical framework and quantifies the expected constraining power under various observational scenarios. Future work will apply these tools to existing spectroscopic datasets including LAMOST, DESI, and other surveys to derive actual constraints on the MW-LMC system. Our publicly available simulation suite will facilitate broader applications in MW-LMC studies.

This paper is organized as follows: Section \ref{sec:simulations} describes our N-body simulation grid construction. Section \ref{sec:results} presents the LMC-induced dynamical effects and our inference framework, examining how observational strategies and uncertainties influence parameter constraints through Fisher matrix forecasts. Section \ref{sec:discussion} addresses systematic uncertainties, survey implications, and current limitations. Section \ref{sec:conclusions} summarizes our findings.

\section{Simulations}
\label{sec:simulations}

We use fully self-consistent N-body simulations to model the MW-LMC interaction, where both galaxies are represented as live N-body systems. This approach captures non-linear effects like dynamical friction more accurately than analytical approximations. While hydrodynamical simulations would include baryonic processes such as gas physics, star formation, and stellar feedback \citep[e.g.,][]{2017MNRAS.467..179G, 2021ApJ...921L..36L,2023ApJS..265...44W}, dark matter-only N-body simulations suffice for studying the LMC's dynamical influence on the MW halo, as demonstrated by previous work \citep[e.g.,][]{2019ApJ...884...51G, 2021MNRAS.501.2279V, 2024MNRAS.527..437V, 2024A&A...688A..51J}.

Building on our previous work \citep{2024MNRAS.534.2694S}, we construct a grid of 2,848 simulations that systematically explore MW and LMC halo masses and shapes, assuming the LMC is on its first infall. This expanded parameter space enables investigation of the MW halo's dynamical response across diverse interaction scenarios.

The simulation process involves: (1) generating initial galaxy models (Section~\ref{sec:Initial models of galaxies}), (2) defining the parameter space (Section~\ref{sec:parameter space}), (3) reconstructing the LMC's initial orbit (Section~\ref{sec:orbital reconstruction}), (4) running high-resolution simulations (Section~\ref{sec:high-resolution}), and (5) creating mock stellar halos (Section~\ref{sec:stellar_tagging}).

\subsection{Initial models of galaxies}
\label{sec:Initial models of galaxies}

We construct initial N-body models of the MW and LMC using the \textsc{galic} code \citep{2014MNRAS.444...62Y}, which generates equilibrium galaxy models based on density distributions and velocity anisotropy profiles.

Our Milky Way models follow the structure of MWPotential2014 from \citet{2015ApJS..216...29B}, consisting of three components: a dark matter halo, stellar disk, and stellar bulge. The dark matter halo follows a Navarro-Frenk-White (NFW) profile \citep{1997ApJ...490..493N} with variable virial mass $M_{200}$, concentration parameter $c = r_{200}/r_s$, and flattening parameter $q$ (which characterizes the halo's axis ratio perpendicular to the Galactic disk).

The stellar disk uses a Miyamoto-Nagai profile \citep{1975PASJ...27..533M} with mass $M_d = 6.8 \times 10^{10} {\rm M}_\odot$, scale length $a = 3$ kpc, and scale height $b = 0.28$ kpc. The bulge follows a Hernquist profile \citep{1990ApJ...356..359H} with mass $M_b = 0.5 \times 10^{10} {\rm M}_\odot$ and scale radius $a_h = 0.54$ kpc. These disk and bulge components remain fixed across all models to ensure stable inner potentials and accurate center-of-mass tracking, while maintaining consistency with stellar mass–halo mass relations for MW-like galaxies \citep[e.g.,][]{2019MNRAS.488.3143B}.

The LMC is modeled as a spherical Hernquist dark matter halo. Though the actual LMC contains a stellar disk, we omit this component since the dominant dynamical perturbations affecting the MW halo arise from the LMC's dark matter halo \citep{2002AJ....124.2639V}. The Hernquist scale radius $a_{h}$ is determined by matching the circular velocity at $r = 8.7$ kpc to the observed value of $\sim$92 km/s \citep{2014ApJ...781..121V}:
\begin{equation}
v_c(r=8.7\mathrm{~kpc})=\frac{\sqrt{G M_{200} r}}{r+a_h}\approx 92 \mathrm{~km/s}
\end{equation}

Before combining the systems, we evolve each galaxy in isolation for 3 Gyr using \textsc{gadget-4} \citep{2021MNRAS.506.2871S} to ensure dynamical stability. Following \citet{2019ApJ...884...51G}, who demonstrated that isolated MW models show variations of less than 5\% in density and velocity anisotropy profiles after 2.5 Gyr, our 3 Gyr relaxation ensures transient disequilibria are sufficiently damped before the interaction begins.

\subsection{Explored Parameter Space}
\label{sec:parameter space}

\renewcommand{\arraystretch}{1.3}
\setlength{\tabcolsep}{4.pt}
\begin{table}
    \centering
    \caption{Summary of the explored MW–LMC parameter space. The four parameters shown are varied using Latin Hypercube Sampling to ensure efficient and uniform coverage of the multidimensional space. We consider two representative assumptions for the MW halo velocity anisotropy: (1) isotropic ($\beta(r) = 0$) and (2) radially varying ($\beta(r) = -0.15 - 0.2\alpha(r)$). Our fiducial model adopts $M_{\mathrm{MW}} = 0.7 \times 10^{12} {\rm M}_\odot$, $M_{\mathrm{LMC}} = 1.5 \times 10^{11} {\rm M}_\odot$, $c = 9.415$, $q = 1.0$, and an isotropic velocity profile.}
	\label{tab:parameter_table}
	\begin{tabular}{cccc} 
		\hline
		$M_{\mathrm{MW}}$ ($10^{12}{\rm M}_{\odot}$) & $M_{\mathrm{LMC}}$ ($10^{11}{\rm M}_{\odot}$) & $c$ & $q$ \\
		\hline
		$0.5-2.0$ & $0.7-2.1$ & $5-15$ & $0.5-1.5$\\
		\hline
	\end{tabular}
\end{table}

We explore a four-dimensional parameter space defined by MW and LMC halo properties. For the MW, we vary the halo virial mass $M_{\mathrm{MW}}$ from $0.5 \times 10^{12} {\rm M}_\odot$ to $2.0 \times 10^{12} {\rm M}_\odot$, covering recent observational and theoretical estimates \citep{2020SCPMA..6309801W,2022ApJ...925....1S}. The concentration parameter $c$ ranges from 5 to 15, accounting for scatter in mass-concentration relations. The halo shape parameter $q$ varies from 0.5 (oblate) to 1.5 (prolate), encompassing plausible halo shapes.

For the LMC, we vary the halo virial mass $M_{\mathrm{LMC}}$ from $0.7 \times 10^{11} {\rm M}_\odot$ to $2.1 \times 10^{11} {\rm M}_\odot$. The lower bound represents the minimum mass supporting the observed LMC stellar disk \citep{2019ApJ...884...51G}, while the upper bound reflects timing argument estimates \citep{2016MNRAS.456L..54P}. Table~\ref{tab:parameter_table} summarizes the parameter space.

We employ Latin Hypercube Sampling (LHS) to generate 5,000 distinct parameter combinations. LHS ensures uniform coverage of each parameter range while minimizing clustering in high-dimensional space, providing more complete and efficient coverage than random sampling \citep{ef76b040-2f28-37ba-b0c4-02ed99573416}.

The velocity anisotropy parameter $\beta(r)$ characterizes the degree of radial bias in stellar orbits \citep{2008gady.book.....B}\footnotemark:
\begin{equation}
\beta \equiv 1-\frac{\sigma_{v_\phi}^2+\sigma_{v_\theta}^2}{2\sigma_{v_r}^2} 
\end{equation}

\footnotetext{Throughout this paper, $v_l \equiv v_\phi$ and $v_b \equiv v_\theta$, i.e. the Galactocentric longitudinal and latitudinal velocity components of stars.}

Cosmological simulations predict radially-biased orbits ($\beta(r) > 0$) for MW stellar halos \citep[e.g.,][]{2008ApJ...681.1076D,2017MNRAS.464.2882A,2018ApJ...853..196L}. Gaia-based studies measure $\beta \simeq 0.8$–0.9 within 25–30 kpc \citep{2019AJ....157..104B,2021MNRAS.502.5686I}, but constraints beyond this radius remain sparse \citep[e.g.,][]{2024arXiv240601676C}.

Since our focus is the outer halo ($R_{\mathrm{gal}} > 30$ kpc) where $\beta(r)$ remains uncertain, we adopt two representative scenarios rather than treating it as a free parameter: (1) isotropic with $\beta(r) = 0$, and (2) radially varying following \citet{2006NewA...11..333H}:
\begin{equation}
\beta(r) = -0.15 - 0.2 \alpha(r), \quad \text{with} \quad \alpha(r) = \frac{\mathrm{d} \ln \rho(r)}{\mathrm{d} \ln r}.
\end{equation}
These profiles bracket the plausible range and allow us to assess the robustness of our results without overfitting uncertain quantities.

We focus on first-infall models, retaining only simulations where the LMC completed only one pericentric passage and reached an apocenter beyond the MW's virial radius within the last 5 Gyr. We adopt first-infall models for two reasons. First, recent observational constraints \citep[e.g.,][]{2013ApJ...764..161K,2020ApJ...893..121P}, including our previous study \citep{2024MNRAS.534.2694S}, support this scenario. Our work demonstrated that the amplitude of the LMC's dynamical effects on MW halo stars favors a massive LMC currently at first infall, as a second infall would induce a weaker LMC wake than has been observed.

Second, focusing on first-infall models is a practical necessity for systematic parameter exploration. A key challenge in previous studies has been orbital reconstruction for two-body N-body systems. At the parameter boundary between first-infall and second-infall scenarios, a drastic shift occurs in the orbital reconstruction, making systematic parameter space exploration considerably more difficult. By restricting to first-infall models, we ensure smooth parameter space coverage and robust orbital solutions across our simulation grid.

After applying this criterion, we retain 1,848 parameter combinations for $\beta(r) = 0$. We randomly select 1,000 of these to simulate with $\beta(r) = -0.15 - 0.2\alpha(r)$ for comparison, yielding our total of 2,848 simulations.

\begin{figure*}
	\includegraphics[width=2.0\columnwidth]{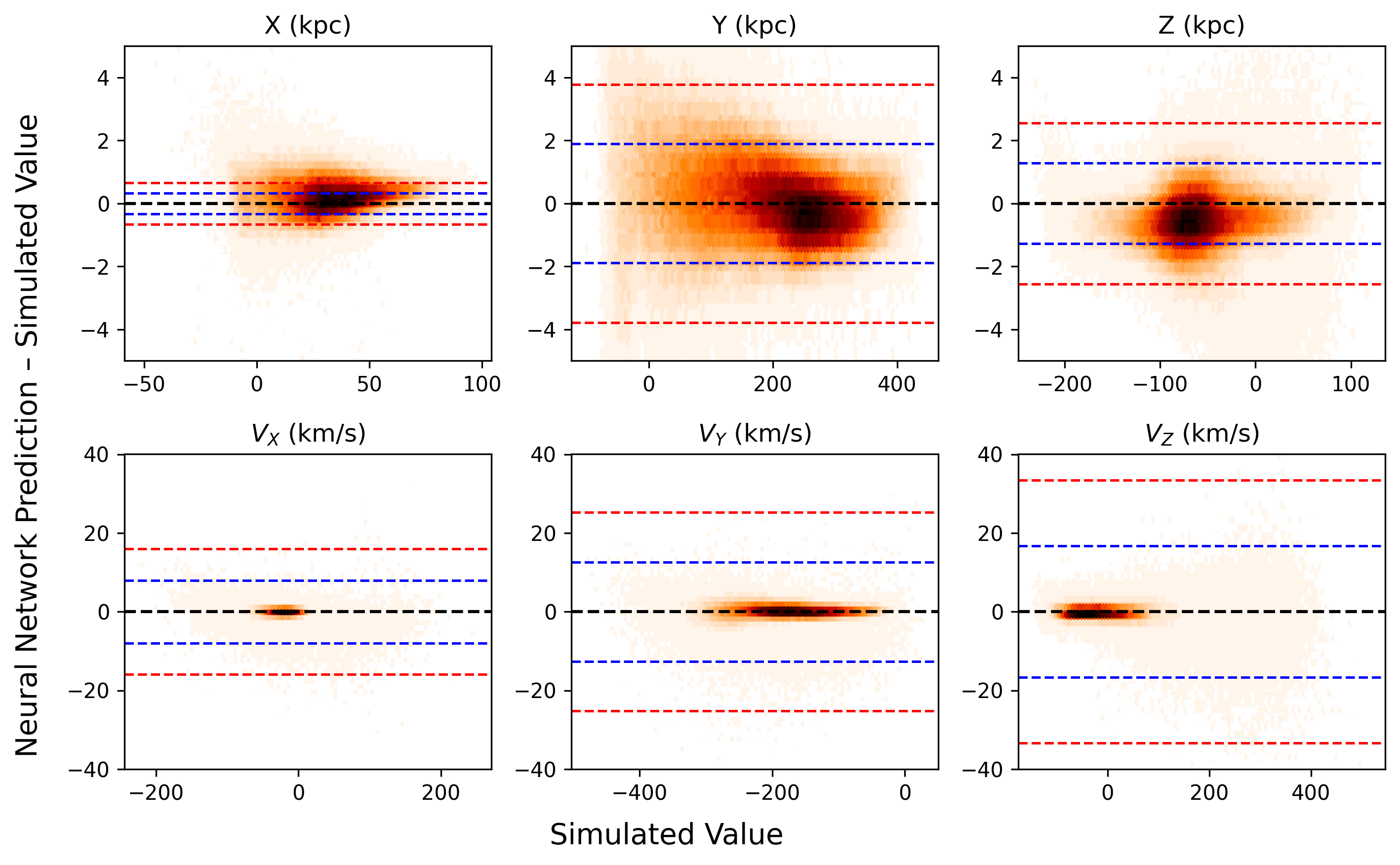}
    \caption{Residuals between the neural network predictions and the simulated phase-space coordinates of the LMC relative to the MW. Each panel corresponds to one of the six phase-space dimensions: position ($X$, $Y$, $Z$ in kpc) in the top row and velocity ($V_X$, $V_Y$, $V_Z$ in km/s) in the bottom row. The horizontal axis shows the simulated value from low-resolution N-body runs, while the vertical axis shows the difference between the predicted and simulated values (i.e., prediction error). Each hexbin map displays the density of samples in the residual space. The black dashed line indicates perfect predictions, while the blue and red dashed lines mark the $\pm1\sigma$ and $\pm2\sigma$ current observational uncertainties of the LMC phase space information, respectively. The residuals in the position coordinates predominantly lie within the $2\sigma$ bounds, while the velocity residuals are tightly clustered within the $1\sigma$ level, indicating that the neural network provides accurate predictions for both position and velocity, suitable for reconstructing the orbital history of the LMC at different physical parameters.}
    \label{fig:model_compare}
\end{figure*}

\subsection{Orbital reconstruction}
\label{sec:orbital reconstruction}

The initial conditions described above define independent N-body realizations for the MW and LMC. To initialize each interaction simulation, we must determine the LMC's relative position and velocity when it first enters the MW's virial radius, such that its evolution leads to the observed present-day phase-space coordinates. This requires reconstructing the LMC's past orbit for each MW–LMC parameter combination. 

The present-day phase-space coordinates of the LMC relative to the MW are taken from \citet{2013ApJ...764..161K}, converted to the Galactocentric Cartesian frame by adopting a solar circular velocity of $V_{\mathrm{c,peak}}(8.29\mathrm{kpc}) \approx 239\mathrm{km/s}$ \citep{2011MNRAS.414.2446M}, and solar peculiar velocity $(U, V, W)_{\odot} = (11.1^{+0.69}_{-0.75}, 12.24^{+0.47}_{-0.47}, 7.25^{+0.37}_{-0.36})\mathrm{km/s}$ \citep{2010MNRAS.403.1829S}. After propagating observational uncertainties through Monte Carlo sampling of the measured proper motions, radial velocities, and distance, we obtain the following present-day Galactocentric coordinates:
$X = -1.06 \pm 0.33$ kpc,
$Y = -41.05 \pm 1.89$ kpc,
$Z = -27.83 \pm 1.28$ kpc,
$V_x = -57.60 \pm 7.99$ km/s,
$V_y = -225.96 \pm 12.60$ km/s, and
$V_z = 221.16 \pm 16.68$ km/s.
These values serve as observational constraints in our reconstruction framework.

If the MW and LMC were point sources, backward orbital integration would suffice. However, both are deformable N-body systems. The LMC's substantial mass (10-20\% of the MW's) displaces the MW's disk from the Galactic center, creating a time-varying potential. Additionally, dynamical friction on the LMC exceeds predictions from classical Chandrasekhar theory \citep{1943ApJ....97..255C}. These effects make backward integration in fixed potentials unreliable for determining initial conditions.

Previous studies \citep[e.g.,][]{2019ApJ...884...51G,2021MNRAS.501.2279V,2021ApJ...921L..36L,2024MNRAS.534.2694S} used iterative refinement for orbital reconstruction, starting from approximate backward integration and refining through multiple rounds of low-resolution N-body simulations. While more accurate, this approach becomes computationally prohibitive for thousands of parameter combinations.

We develop an inverse modeling approach using a feedforward Multi-Layer Perceptron (MLP) neural network trained to map initial phase-space coordinates $(\mathbf{x}_0, \mathbf{v}_0)$, model parameters $\boldsymbol{\theta} = (M_{\mathrm{MW}}, M_{\mathrm{LMC}}, c, q)$, and time $T$ to the LMC's evolved coordinates:
\begin{equation}
\mathbf{x}_{\mathrm{pred}}(T), \mathbf{v}_{\mathrm{pred}}(T) = f_{\mathrm{MLP}}(\mathbf{x}_0, \mathbf{v}_0, T, \boldsymbol{\theta}).
\end{equation}

The network architecture consists of an initial linear layer mapping 11-dimensional input (6D initial phase-space coordinates, 4D model parameters, and 1D time), eight hidden layers (the first hidden layer contains 64 neurons, and the rest 128 neurons each with ReLU activation), and a 6D output layer. Training uses the Adam optimizer (learning rate 0.001), cosine-annealing scheduler, mean-squared error loss, 80:20 train-test split, and batch size 256.

Training data comprises seven forward-simulated orbits per parameter combination. We first obtain approximate initial conditions using analytical backward integration with \texttt{galpy} \citep{2015ApJS..216...29B}. Though not accurate for N-body systems, these provide starting points around which we generate six additional perturbed orbits. This strategy efficiently covers the relevant phase space while maintaining physical motivation. All training uses low-resolution simulations ($\sim10^6$ particles).

\begin{figure*}
	\includegraphics[width=2.0\columnwidth]{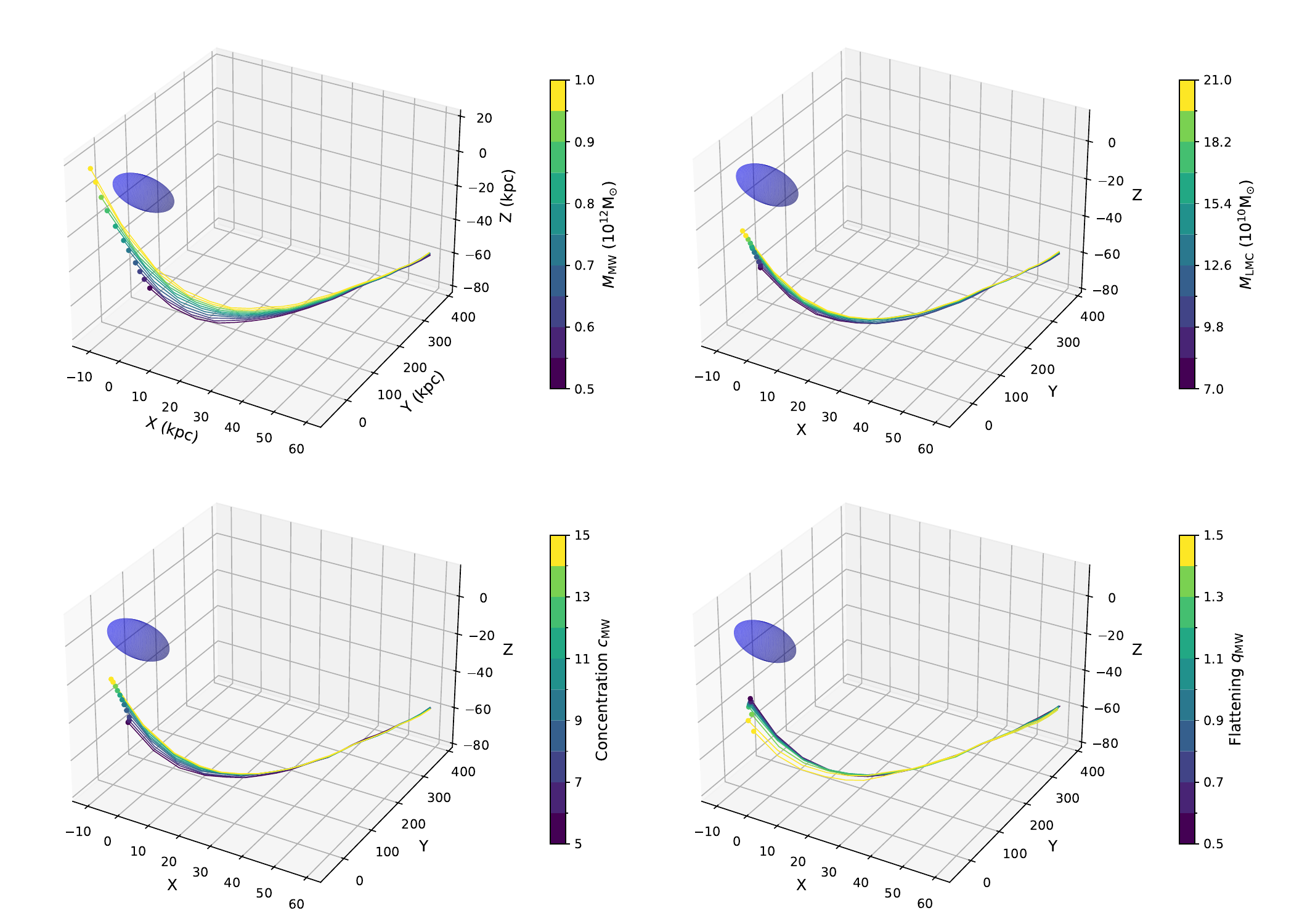}
    \caption{Predicted past trajectories of the LMC relative to the Galactic disk (shown as a blue circle in the x–y plane at $z=0$), generated by our trained neural network under the first-infall orbital scenario. Each orbit starts from the same initial phase-space coordinates, and the trajectories are evolved under different MW–LMC model parameters. The fiducial model adopts $M_{\text{MW}} = 0.7 \times 10^{12} \mathrm{M}_{\odot}$, $M_{\text{LMC}} = 1.5 \times 10^{11} \mathrm{M}_{\odot}$, concentration $c = 9.415$, and halo flattening $q = 1.0$. Each panel varies one parameter at a time while keeping the others fixed: MW halo mass (upper left), LMC mass (upper right), MW concentration (lower left), and halo flattening (lower right). As expected, increasing the MW mass produces a more tightly bound LMC orbit due to a deeper gravitational potential, while increasing the LMC mass leads to greater energy loss from dynamical friction, also tightening the orbit. The smooth evolution of trajectories with varying parameters ensures that the resulting dynamical perturbations in the MW halo also vary smoothly across parameter space, enabling robust emulation of kinematic observables.}
    \label{fig:LMC_orbits}
\end{figure*}

Figure~\ref{fig:model_compare} shows residuals between neural network predictions and simulation outputs. Predicted positions lie within $2\sigma$ of observational uncertainties, while velocities lie within $1\sigma$, demonstrating sufficient precision for orbital reconstruction. Figure \ref{fig:LMC_orbits} illustrates how predicted trajectories vary with MW–LMC parameters from identical initial conditions. Using a fiducial model ($M_{\text{MW}}=0.7 \times 10^{12} \mathrm{M}_{\odot}$, $M_{\text{LMC}}=1.5 \times 10^{11} \mathrm{M}_{\odot}$, $c = 9.415$, $q = 1.0$, $\beta(r)=0$), we vary each parameter independently. The resulting orbits evolve smoothly with parameter changes, consistent with theoretical expectations \citep{2023Galax..11...59V}: increasing MW mass deepens the potential well, producing tighter orbits; increasing LMC mass enhances dynamical friction, also tightening orbits.

For any set of MW-LMC parameter $\theta$, we perform optimization to infer initial coordinates $\mathbf{x}_0$ that reproduce observed present-day coordinates after 2 Gyr evolution:
\begin{equation}
\min _{\mathbf{x}_0} \mathcal{L}\left(\mathbf{x}_0\right)=\min _{\mathbf{x}_0}\left\|\frac{f_{\mathrm{MLP}}\left(\mathbf{x}_0, T=2\mathrm{Gyr}, \boldsymbol{\theta}\right)-\mathbf{x}_{\text {obs }}(T=2\mathrm{Gyr})}{\boldsymbol{\sigma}_{\mathrm{obs}}}\right\|^2.
\end{equation}

The 2 Gyr integration time ensures initial coordinates lie outside the MW's virial radius for first-infall scenarios. The loss function weights residuals by observational uncertainties, appropriately scaling each component's contribution. We employ the \texttt{Nelder-Mead} method from \texttt{scipy.optimize} for minimization.

\subsection{Study perturbations in high-resolution simulations}
\label{sec:high-resolution}

With the initial orbital coordinates determined for each MW-LMC parameter combination, we proceed to high-resolution simulations to capture the LMC-induced dynamical perturbations in the MW halo. The low-resolution simulations used for orbital reconstruction lack sufficient particle density to resolve small-scale structures such as the dynamical friction wake and local density perturbations, which occur on scales of several kiloparsecs.

The resolution requirements for capturing these perturbations depend on the number of particles, as demonstrated by \citet{2007MNRAS.375..425W}. Our high-resolution simulations employ $\sim10^{7}$ particles total, with a particle mass of $1 \times 10^5 \mathrm{M}_{\odot}$ for both MW and LMC models. This resolution has been validated in previous studies of MW-LMC interactions \citep[e.g.,][]{2009MNRAS.400.1247C,2019ApJ...884...51G,2024MNRAS.534.2694S} as sufficient to resolve the dynamical features relevant to our analysis, including the wake structure, reflex motion, and velocity field perturbations in the halo.

The gravitational softening lengths are set following the criteria of \citet{2003MNRAS.338...14P}: $\epsilon=\frac{4 r_{200}}{\sqrt{N_{200}}}$, where $N_{200}$ is the number of particles within the virial radius $r_{200}$ of the MW. This yields softening lengths of 200-240 pc for high-resolution simulations and 600-750 pc for low-resolution runs, ensuring that the force resolution is appropriately scaled to the particle density in each case.

We evolve each MW-LMC system for 2 Gyr using \textsc{gadget-4}, sufficient to capture the present-day configuration after the LMC's infall from beyond the MW's virial radius. The computational requirements scale notably with resolution: low-resolution simulations require approximately 16 CPU hours each, while high-resolution runs demand 600 CPU hours. Given our suite of 2,848 parameter combinations for the high-resolution runs, the total computational effort exceeds 1.8 million CPU hours.

\subsection{Constructing the Stellar Halo}
\label{sec:stellar_tagging}

\begin{figure*}
	\includegraphics[width=2\columnwidth]{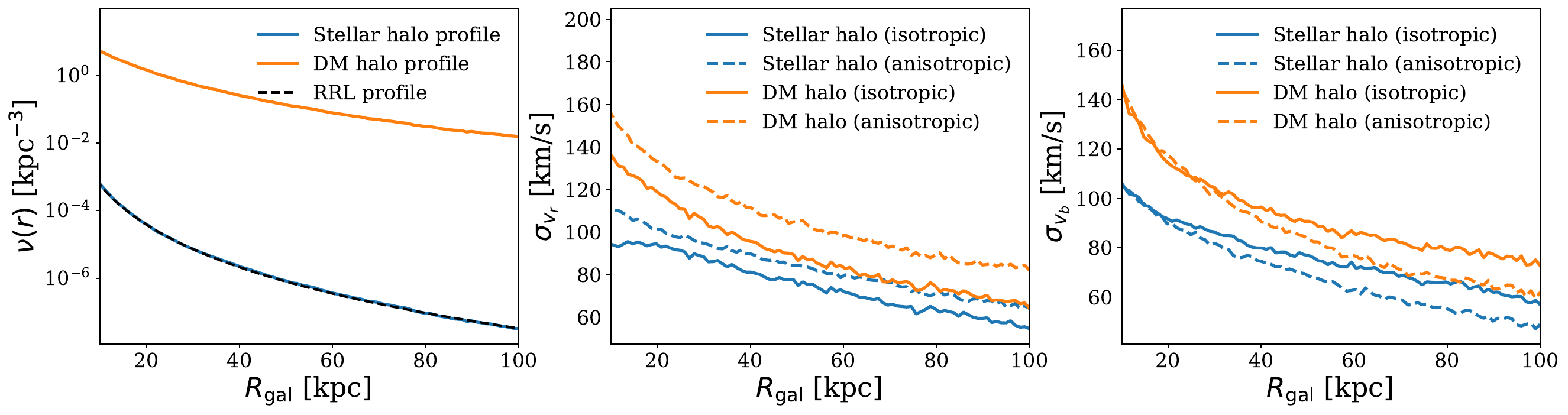}
    \caption{Initial properties of the mock stellar halo constructed by tagging dark matter (DM) particles in the isolated, equilibrium fiducial MW model ($M_{\text{MW}}=0.7\times10^{12}\mathrm{M}_{\odot}$, $c$ = 9.415, $q$ = 1.0) prior to LMC's infall. \textbf{Left:} Number density profile of the stellar halo (blue curve) compared with the analytical Einasto profile used in the tagging process (black dashed line), and the underlying DM halo profile (orange curve). \textbf{Middle:} Galactocentric radial velocity dispersion. \textbf{Right:} Galactocentric latitudinal velocity dispersion. Solid curves correspond to an isotropic velocity profile ($\beta=0$), while dashed curves represent a radially varying anisotropic profile ($\beta(r) = -0.15 - 0.2 \alpha(r)$, with $\alpha(r) = \mathrm{d} \ln \rho / \mathrm{d} \ln r$). Across all panels, the mock stellar halo is constructed using a weighting scheme based on Eddington's inversion formula applied to the equilibrium MW model, spanning from 10 to 100 kpc. The stellar halo density aligns with the Einasto profile for RR Lyrae stars. Velocity dispersions for the stellar halo are $\sim$20 km/s lower than the DM halo beyond 20 kpc, with anisotropy raising radial dispersion and lowering tangential dispersion.}
    \label{fig:weight_profiles}
\end{figure*}

Our simulations model the MW and LMC as collisionless dark matter systems, but observational constraints come from the MW's stellar halo. The stellar and dark matter halos differ in their spatial distributions and kinematics: the stellar halo is more centrally concentrated, less massive, and dynamically colder. To compare simulations with observations, we construct a mock stellar halo within the dark matter halo.

Since the stellar halo's self-gravity is negligible compared to the dark matter halo, we adopt an energy-based particle-tagging approach \citep{2013MNRAS.435..901L}. Each dark matter particle receives a weight determining its probability of representing a halo star, allowing us to reproduce the observed stellar density and kinematics. This method offers two key advantages over including a live stellar component: (i) it allows us to construct arbitrary stellar halos in post-processing, providing far higher effective resolution at much lower computational cost - a benefit given the thousands of simulations required in this work; and (ii) it avoids the spurious heating of low-mass stellar particles that often arises in multi-component N-body simulations where massive dark matter particles scatter lighter tracers \citep{2025MNRAS.540.2049C}.

We assign each dark matter particle a weight $\omega(E)$ based on its energy $E$:
\begin{equation}
\omega(E)=\frac{N_\star(E)}{N(E)}=\frac{f_\star(E) g(E)}{N(E)},
\end{equation}
where $N(E)$ and $N_\star(E)$ are the numbers of dark matter and stellar particles at energy $E$, $f_\star(E)$ is the stellar distribution function (phase-space density), and $g(E)$ is the density of states (available phase-space volume). We calculate weights using the isolated MW models before LMC infall, as Eddington's formula requires dynamical equilibrium.

The stellar distribution function follows Eddington's formula:
\begin{equation}
f_\star(\mathcal{E})=\frac{1}{\sqrt{8} \pi^2}\left[\int_0^{\mathcal{E}} \frac{\mathrm{d} \Psi}{\sqrt{\mathcal{E}-\Psi}} \frac{\mathrm{d}^2 \nu(\Psi)}{\mathrm{d} \Psi^2}\right]+\left.\frac{1}{\sqrt{\mathcal{E}}} \frac{\mathrm{~d} \nu}{\mathrm{~d} \Psi}\right|_{\Psi=0},
\label{eq:distribution function}
\end{equation}
which provides the unique isotropic distribution function for a given density profile $\nu(r)$ in potential $\Phi(r)$. Here, $\Psi=-\Phi+\Phi_0$ is the relative potential with $\Phi_0=0$ (since our NFW potential vanishes at infinity), and $\mathcal{E}=\Psi-\frac{v^2}{2}$ is the relative energy.

We adopt the RR Lyrae density profile from \citet{2018ApJ...859...31H}, following an Einasto form:
\begin{equation}
\nu(r)=\nu_e \exp \left\{-d_n\left[\left(\frac{r}{r_{\text {eff }}}\right)^{1 / n}-1\right]\right\},
\end{equation}
with shape parameter $n=9.53$, scale radius $r_{\text {eff }}=1.07$ kpc, and $d_n=3 n-\frac{1}{3}+\frac{0.0079}{n}$ \citep{2006AJ....132.2685M}. The normalization $\nu_e$ sets the total stellar halo mass but does not affect our kinematic analyses.

The density of states is:
\begin{equation}
g(E)=(4 \pi)^2 \int_0^{r_E} r^2 \sqrt{2[E-\Phi(r)]} \mathrm{d} r
\label{eq:density of states}
\end{equation}

For non-spherical MW models (oblate or prolate), we use the spherically averaged potential $\langle \Phi(r) \rangle$ when computing $f_\star(E)$, as Eddington's formula requires monotonic $\Phi(r)$. However, we use actual potential values when calculating particle energies, ensuring the stellar halo reflects the underlying potential shape. This approximation remains valid for mildly flattened potentials \citep{1997MNRAS.286..329N}.

After assigning weights $\omega_i$ to dark matter particles, we compute stellar halo kinematics as weighted averages:
\begin{equation}
\bar{v_{\star}}=\frac{\sum_i \omega_i v_i}{\sum_i \omega_i}
\end{equation}
\begin{equation}
\sigma_{v_{\star}}=\sqrt{\frac{\sum_i \omega_i\left(v_i-\bar{v}\right)^2}{\sum_i \omega_i}}
\end{equation}
where $v_i$ and $\bar{v}$ are the velocity of each DM particle and the mean velocity of DM particles respectively. This ensures the mock stellar halo maintains consistency with the imposed energy distribution while evolving with the LMC-induced perturbations.

Figure~\ref{fig:weight_profiles} shows the initial mock stellar halo properties for our fiducial MW model before LMC infall. The stellar density matches the adopted Einasto profile, while velocity dispersions are $\sim$20 km/s lower than the dark matter halo beyond 20 kpc, reflecting the colder stellar kinematics. The anisotropic velocity profile ($\beta(r) = -0.15 - 0.2\alpha(r)$) increases radial dispersion while decreasing tangential dispersion compared to the isotropic case. The tagged stellar halo then evolves self-consistently throughout the MW-LMC interaction.

\begin{figure*}
	\includegraphics[width=2.0\columnwidth]{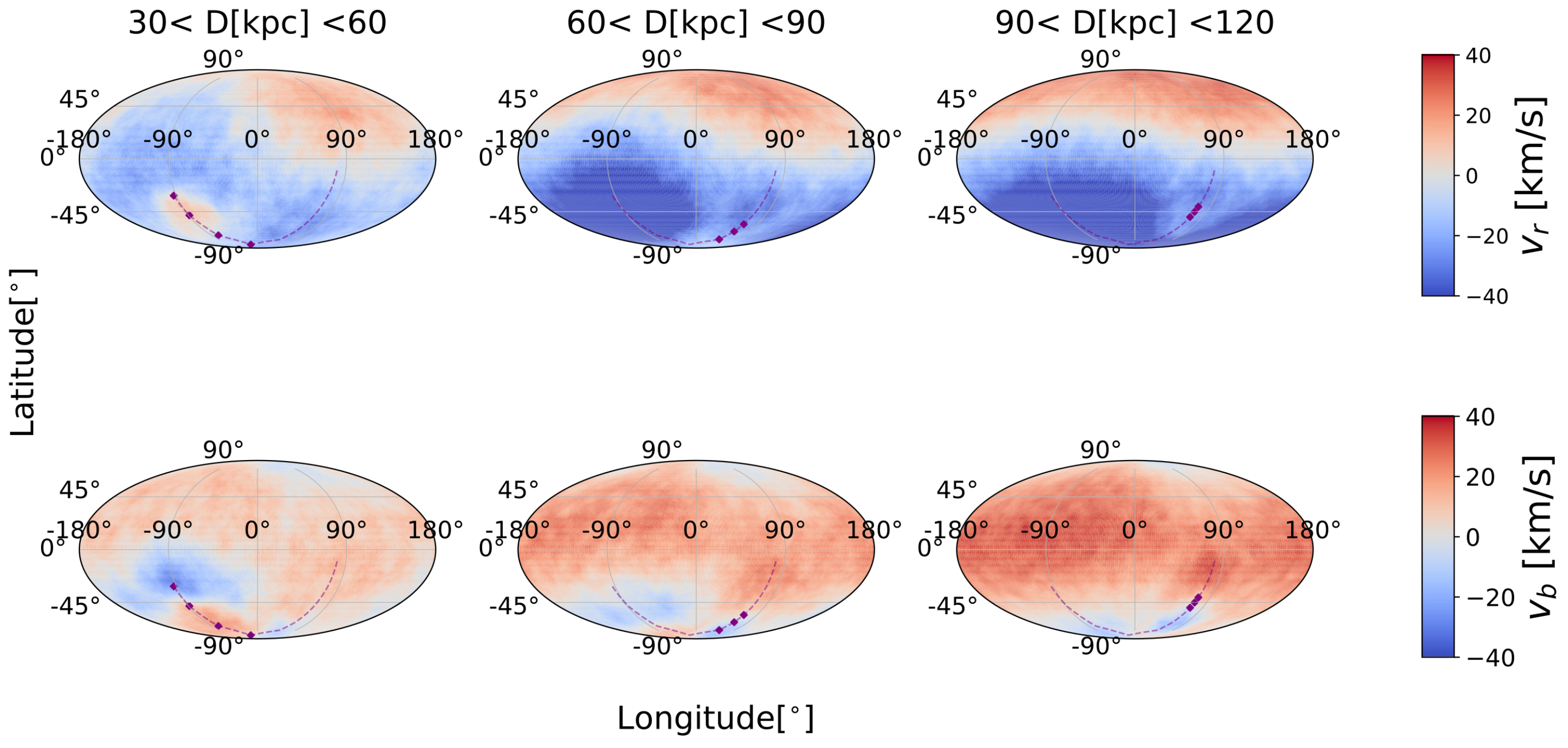}
    \caption{Galactocentric sky maps of kinematic perturbations in the fiducial case ($M_{\text{MW}}=0.7\times10^{12}\mathrm{M}_{\odot}$, $M_{\text{LMC}}=1.5\times10^{11}\mathrm{M}_{\odot}$, $c=9.415$, $q=1.0$, isotropic velocity profile $\beta(r)=0$). We show the line-of-sight velocity (top) and latitudinal velocity (bottom) maps for halo stars in three radial ranges: 30–60 kpc (left), 60–90 kpc (middle), and 90–120 kpc (right). The LMC trajectory is marked with a dashed purple line, with solid diamonds indicating positions within specific radial ranges. Since the initial conditions of the MW halo in our N-body simulations assume equilibrium, the mean velocities of the stars are close to zero prior to the infall of the LMC. A disparity in dynamical timescales leads to a differential response between the inner and outer halo, creating a north-south dipole asymmetry in radial velocity and a global positive bias in latitudinal velocity.}
    \label{fig:fiducial_mean_map}
\end{figure*}

\begin{figure*}
	\includegraphics[width=2.0\columnwidth]{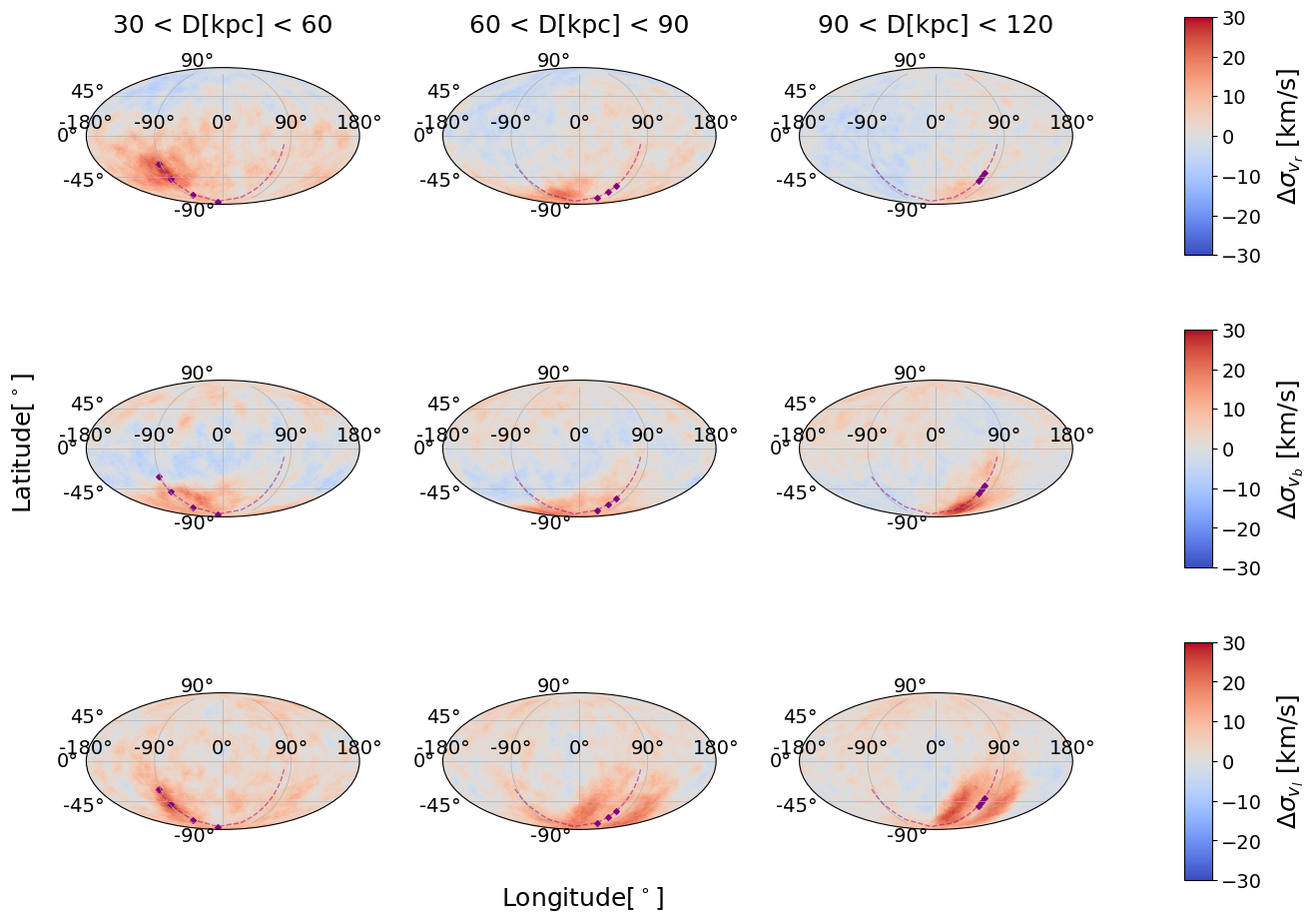}
    \caption{Similar to Figure \ref{fig:fiducial_mean_map}, this figure shows Galactocentric sky maps of the changes in velocity dispersion for halo stars following the LMC's infall in the fiducial MW–LMC simulation. The rows correspond to different velocity components: radial ($v_r$, top), latitudinal ($v_b$, middle), and longitudinal ($v_l$, bottom), while the columns represent Galactocentric distance bins of 30–60 kpc (left), 60–90 kpc (center), and 90–120 kpc (right). While localized enhancements are visible along the LMC's trajectory, the majority of the sky exhibits only mild perturbations (typically less than 5 km/s), indicating that the LMC has a limited effect on the second-moment kinematic statistics of stars. This contrasts with the pronounced shifts observed in the mean velocities of halo stars, which reflect the coherent kinematic response to the recent perturbations induced by the LMC. In comparison, velocity dispersions are largely governed by the equilibrium structure of the MW halo. This distinction motivates the combined use of both mean velocity and velocity dispersion maps to disentangle degeneracies in the MW–LMC parameter space.}
    \label{fig:fiducial_std_map}
\end{figure*}

\section{Results}
\label{sec:results}

We investigate how stellar halo kinematics encode information about the MW–LMC interaction and assess their effectiveness in constraining model parameters. We characterize the kinematic signatures using mean velocity and velocity dispersion of halo stars across three Galactocentric radial ranges: 30–60 kpc, 60–90 kpc, and 90–120 kpc. These bins are wider than typical distance uncertainties and broad enough to minimize sampling noise.

Section~\ref{sec:dynamical effects} builds physical intuition from simulations, showing that mean velocity (first moment) and velocity dispersion (second moment) respond differently to LMC-induced perturbations. This understanding motivates using both statistics in our inference framework. Section \ref{sec:posterior_inference} derives posterior distributions of MW–LMC parameters from simulated observations. Section \ref{sec:forecast} presents Fisher matrix forecasts evaluating how observational uncertainties affect parameter inference.

\subsection{Complementary Insights from First and Second Moments}
\label{sec:dynamical effects}

Before quantitative inference, we develop intuition for how mean velocity and velocity dispersion respond to LMC perturbations using our fiducial simulation. This understanding clarifies why these observables are complementary for constraining MW–LMC parameters.

Figures \ref{fig:fiducial_mean_map} and \ref{fig:fiducial_std_map} show Galactocentric sky maps of kinematic perturbations for our fiducial case: $M_{\text{MW}}=0.7\times10^{12}\mathrm{M}_{\odot}$, $M_{\text{LMC}}=1.5\times10^{11}\mathrm{M}_{\odot}$, $c=9.415$, $q=1.0$, and isotropic velocity profile ($\beta(r)=0$). The LMC trajectory appears as a dashed purple line with solid diamonds marking positions within each radial range.

Figure \ref{fig:fiducial_mean_map} displays line-of-sight velocity ($v_r$, top) and latitudinal velocity ($v_b$, bottom) for three radial bins: 30-60 kpc (left), 60-90 kpc (middle), and 90-120 kpc (right). The MW halo begins in equilibrium with near-zero mean velocities before LMC infall. We exclude the inner 30 kpc where particles move with the Galactic disk and show negligible perturbations.

The differential response arises from varying dynamical timescales across the halo (approximately 0.3, 0.5, and 1.0 Gyr at 30, 50, and 100 kpc respectively). The inner galaxy responds faster to the LMC's passage than the outer halo. As a result, the inner MW accelerates toward the LMC (appearing to move downward relative to the disk), while the outer halo maintains negligible net velocity. From the perspective of an observer in the inner MW, this creates the appearance of the outer halo moving upward in the Galactocentric frame.

\begin{figure}
	\includegraphics[width=\columnwidth]{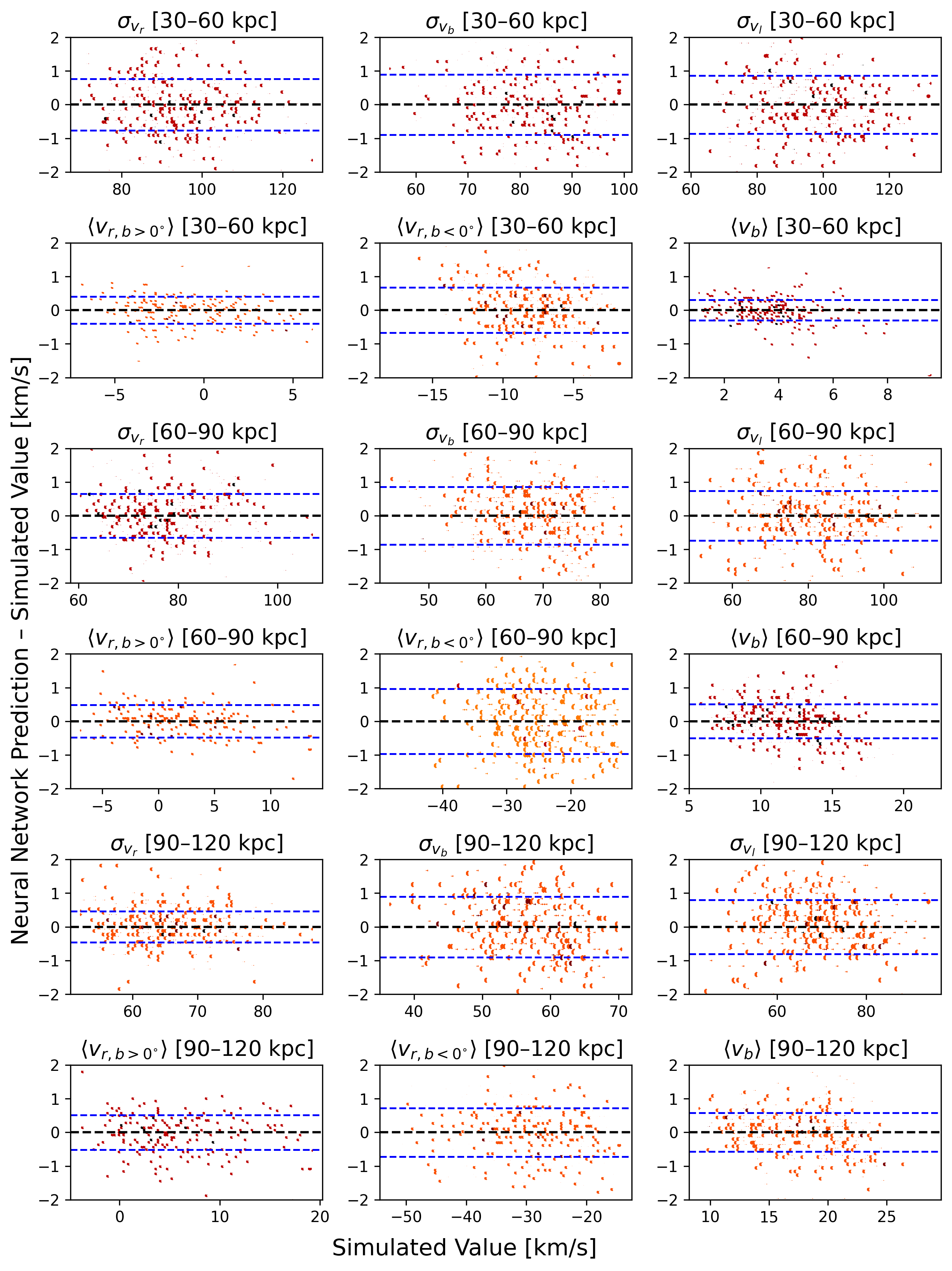}
    \caption{Validation of the neural network emulation for the kinematic summary statistics of halo stars. Each panel shows the residuals (predicted minus simulated values) as a function of the simulated value for the mean velocity ($\langle v_{r,b>0^{\circ}} \rangle$, $\langle v_{r,b<0^{\circ}} \rangle$, $\langle v_b \rangle$) and velocity dispersion ($\sigma_{v_r}$, $\sigma_{v_b}$, $\sigma_{v_l}$) in three Galactocentric distance bins: 30–60 kpc, 60–90 kpc, and 90–120 kpc. The density of points is visualized using hexbin shading. The black dashed line indicates perfect prediction (zero residual), while the blue dashed lines denote the $\pm 1\sigma$ bounds, where $\sigma$ is estimated robustly from the residuals using the median absolute deviation (MAD). For all summary statistics, the typical residual scatter is $\lesssim 1$ km/s—smaller than the observational uncertainty, which includes both individual measurement errors and finite sampling noise within each radial bin. As a result, we neglect the emulation uncertainty in the likelihood function.}
    \label{fig:NN_validation}
\end{figure}

This differential motion produces a north-south dipole in radial velocity: northern stars recede while southern stars approach. The dipole strengthens with distance. At 30-60 kpc, northern radial velocities remain below 10 km/s while southern velocities reach -10 km/s. In the outermost regions (90-120 kpc), southern velocities can reach -30 km/s.

The offset between inner and outer halo creates positive latitudinal velocities across most of the sky. The perturbation amplitude increases with distance: approximately 5 km/s at 30-60 kpc, 15 km/s at 60-90 kpc, and 20 km/s at 90-120 kpc. The LMC minimally affects the longitudinal velocity ($v_l$) in our simulations because its orbital plane is nearly perpendicular to the Galactic disk plane and does not impact the angular momentum of stars in the $v_l$ direction.

Observational studies suggest the MW halo may already possess rotation \citep{2017MNRAS.470.1259D}, potentially established during early galaxy formation. Since the halo's pre-existing rotational velocity remains uncertain and our models assume $\langle v_l \rangle = 0$, the mean longitudinal velocity provides minimal constraining power in our inference framework. When applying this method to real observations, excluding $\langle v_l \rangle$ from the analysis may be prudent to avoid potential systematic biases arising from the uncertain pre-existing halo rotation. 

Figure \ref{fig:fiducial_std_map} shows velocity dispersion changes for $v_r$ (top), $v_b$ (middle), and $v_l$ (bottom) across the same radial ranges. Apart from localized enhancements along the LMC's trajectory, global perturbations remain weak (typically less than 5 km/s). This weak response reflects the different nature of first and second moments: while mean velocities capture the coherent bulk motion induced by the MW's reflex motion, velocity dispersions primarily reflect the MW halo's intrinsic equilibrium structure.

The contrast between Figures \ref{fig:fiducial_mean_map} and \ref{fig:fiducial_std_map} reveals why these statistics provide complementary constraints. First moments capture recent LMC-induced perturbations and bulk kinematic responses, making them sensitive to $M_{\mathrm{LMC}}$. Second moments reflect the MW halo's intrinsic equilibrium structure, providing better constraints on MW properties like $M_{\mathrm{MW}}$, $c$, and $q$. Section~\ref{sec:interpretation} develops an analytical model that explains this behavior quantitatively. This distinction motivates using both statistics to break parameter degeneracies in our inference framework.

\subsection{Posterior Inference from Simulated Kinematic Summary Statistics}
\label{sec:posterior_inference}

\begin{figure*}
	\includegraphics[width=2\columnwidth]{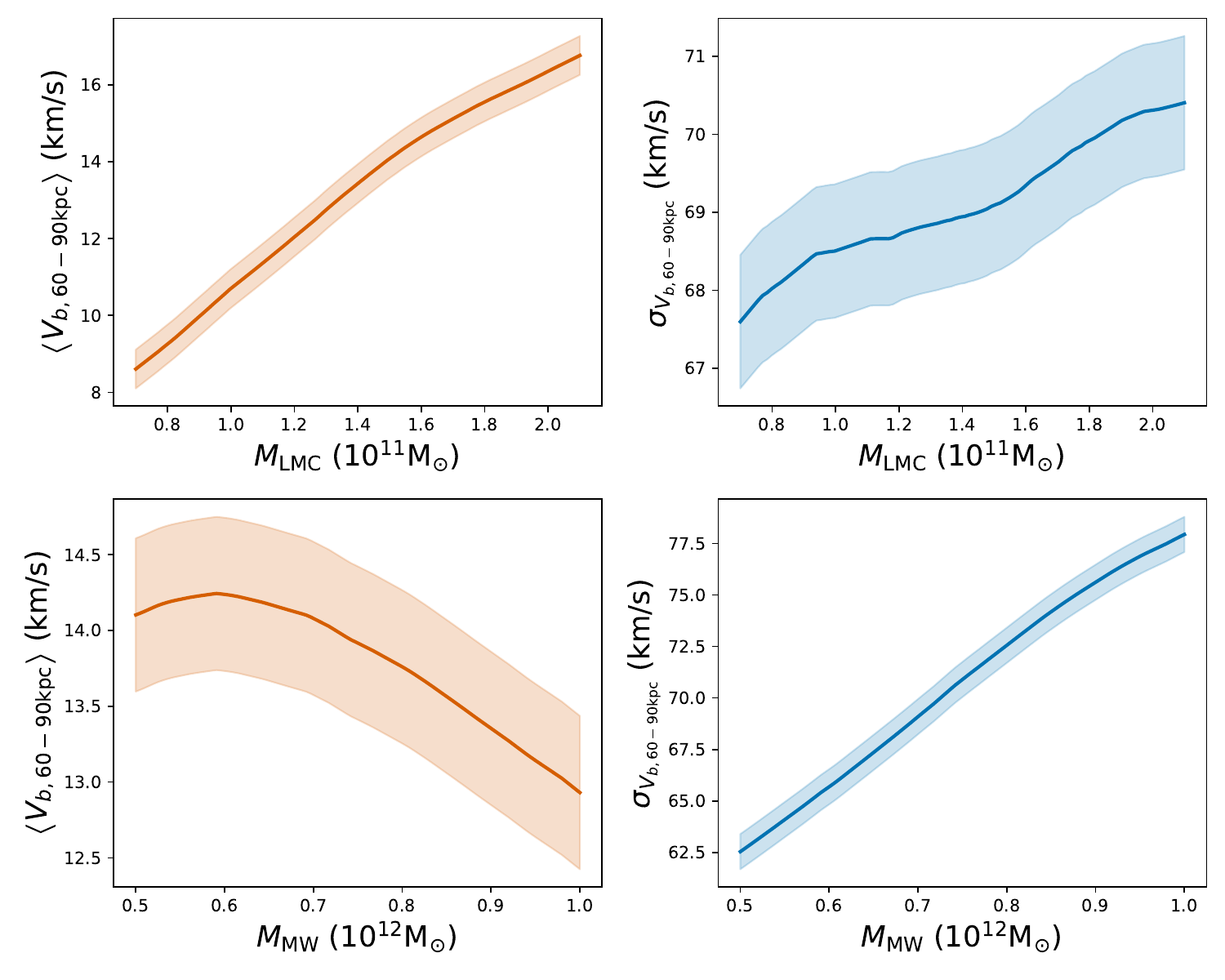}
    \caption{Predicted mean latitudinal velocity ($\langle v_b \rangle$, left) and velocity dispersion ($\sigma_{v_b}$, right) of halo stars at 60–90 kpc, as functions of LMC infall mass ($M_{\mathrm{LMC}}$, top row) and the MW mass ($M_{\mathrm{MW}}$, bottom row), based on neural network emulation. Shaded bands show 1$\sigma$ emulation uncertainties. Other parameters are fixed at fiducial values: $M_{\mathrm{MW}} = 0.7 \times 10^{12}\,{\rm M}_\odot$, $c = 9.415$, $q = 1.0$, and $\beta(r) = 0$. The top row shows that $\langle v_b \rangle$ increases significantly with $M_{\mathrm{LMC}}$, while $\sigma_{v_b}$ shows only a mild change ($\sim$3\%). In contrast, the bottom row shows that increasing $M_{\mathrm{MW}}$ leads to a moderate decrease in $\langle v_b \rangle$ (by roughly 7\%) and a noticeable $\sim24\%$ increase in $\sigma_{v_b}$. These trends indicate that the first moment is more sensitive to perturbations induced by the LMC, whereas the second moment more directly reflects the intrinsic properties of the MW potential.}
    \label{fig:NN_vb_75}
\end{figure*}

\begin{figure*}
	\includegraphics[width=2\columnwidth]{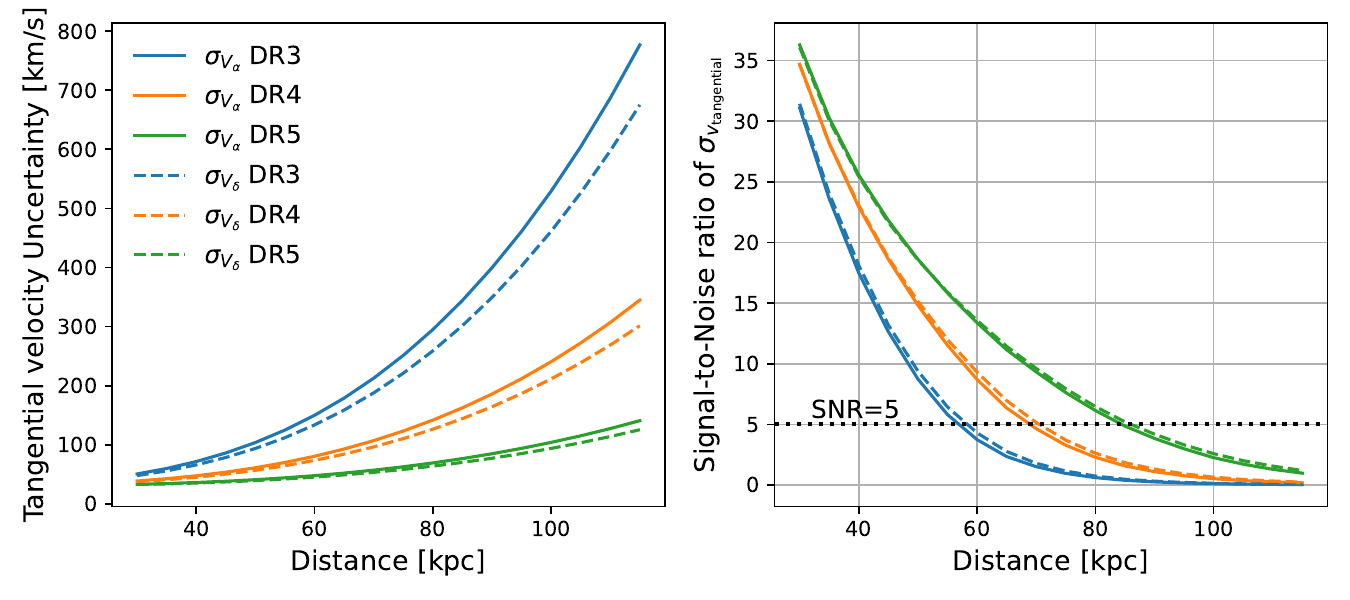}
    	\caption{Measurement precision of individual stellar tangential velocities and the corresponding signal-to-noise ratio (SNR)—defined here as the ratio between the physical tangential velocity dispersion and its statistical sampling uncertainty—as functions of heliocentric distance. \textbf{Left:} Predicted measurement uncertainties in individual tangential velocities of typical halo tracers (e.g., RR Lyrae stars) for Gaia DR3 (blue), DR4 (yellow), and DR5 (green), based on PyGaia estimates scaled for each data release. Beyond $\sim$60 kpc, uncertainties in DR3 rise steeply, exceeding several hundred km s$^{-1}$. \textbf{Right:} Ratio of the stellar halo's physical tangential velocity dispersion ($\sim$60-70 km/s from our fiducial model) to the statistical sampling uncertainty, which scales as $\sim 1/\sqrt{2N}$ where $N$ is the number of stars in each radial bin. Each curve incorporates the fixed measurement precision of its respective Gaia data release. The dashed horizontal line marks the SNR = 5 threshold. For Gaia DR3, the SNR falls below 5 beyond $\sim$60 kpc, while DR4 and DR5 extend this boundary to $\sim$70 and $\sim$90 kpc, respectively. In the outer halo, where tangential velocity dispersion becomes noise-dominated even with Gaia DR5 measurements, radial velocity measurements—typically more precise due to spectroscopy—are crucial for retaining constraining power. Although shown in the heliocentric frame, we use these uncertainty-distance relations as a proxy in the Galactocentric frame, since differences are small at large distances. A full error propagation would be required for real data as it may induce correlations between radial and tangential components.}
    \label{fig:Gaia_uncertainty}
\end{figure*}

\begin{figure}
	\includegraphics[width=1\columnwidth]{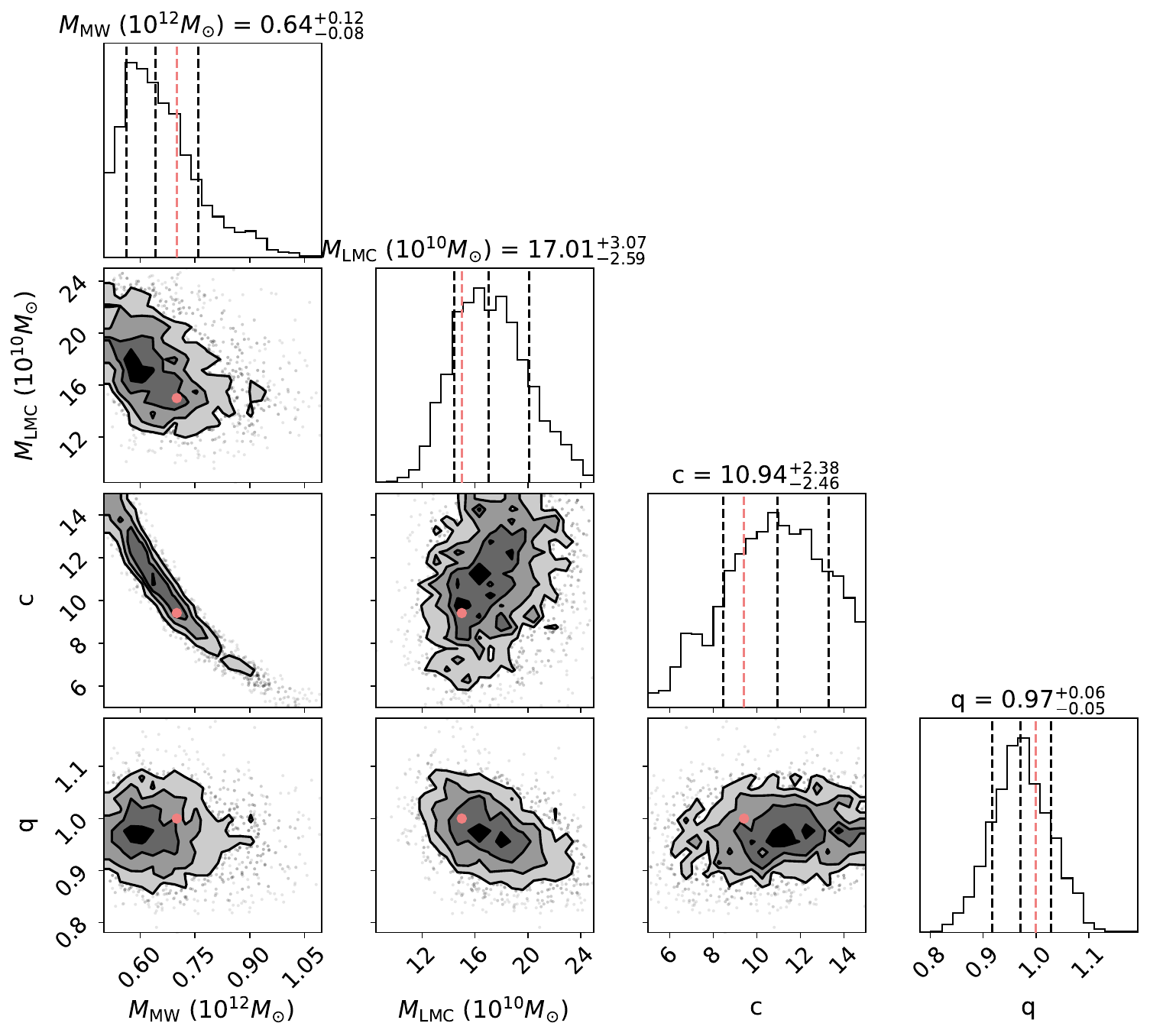}
    \caption{Posterior distributions for $M_{\mathrm{MW}}$, $M_{\mathrm{LMC}}$, $c$, and $q$, inferred from mock observations of the mean velocity and velocity dispersion in the radial, latitudinal, and longitudinal directions for stars within 30–60, 60–90, and 90-120 kpc. The inference is based on summary statistics modeled using neural network emulation trained on simulation outputs, incorporating a fixed 20 km/s measurement uncertainty for stellar radial velocities and Gaia DR3–like uncertainties for tangential velocities. The true parameter values used to generate the mock data ($M_{\mathrm{MW}} = 0.7 \times 10^{12}{\rm M}_\odot$, $M_{\mathrm{LMC}} = 15.0 \times 10^{10}{\rm M}_\odot$, $c = 9.415$, and $q = 1.0$) are indicated by red points in the contour plots and red dashed lines in the diagonal marginal distributions. The marginalized posteriors yield median values and 1$\sigma$ credible intervals of $M_{\mathrm{MW}} = 0.64^{+0.12}_{-0.08} \times 10^{12}{\rm M}_\odot$, $M_{\mathrm{LMC}} = 17.00^{+3.04}_{-2.57} \times 10^{10}{\rm M}_\odot$, $c = 10.91^{+2.38}_{-2.48}$, and $q = 0.97^{+0.06}_{-0.05}$. All parameters are consistent with the true values within the 1–2$\sigma$ regions.}
    \label{fig:posterior_corner}
\end{figure}

The complementary nature of first and second velocity moments motivates their combined use for parameter inference. Since our simulations sample the parameter space discretely, we train a neural network to emulate kinematic summary statistics across the continuous parameter space for first-infall LMC orbits.

We model the mapping:
\[
\boldsymbol{\theta} = (M_{\mathrm{MW}}, M_{\mathrm{LMC}}, c, q) \rightarrow \boldsymbol{y} = (\langle v \rangle, \sigma_v)
\]
where $\boldsymbol{y}$ comprises mean and dispersion of stellar velocities in three Galactocentric bins (30–60, 60–90, and 90–120 kpc) for radial, latitudinal, and longitudinal components. For radial velocity, we separate stars at $b > 0^{\circ}$ and $b < 0^{\circ}$ to capture the dipole pattern from Figure~\ref{fig:fiducial_mean_map}. This yields 18 summary statistics: 3 mean velocities (2 radial components for the northern and southern hemispheres + $v_b$ component) and 3 velocity dispersions per distance bin.

The neural network uses a feedforward architecture with 5 hidden layers: the first contains 64 neurons, the remaining four contain 128 neurons each. We employ ReLU activations, Adam optimizer (learning rate 0.0001), and cosine-annealing scheduler. Training uses an 80:20 split, batch size 256, and mean squared error loss.

Figure \ref{fig:NN_validation} validates the emulation accuracy. Each panel shows residuals (predicted minus simulated) versus simulated values, with density visualized through hexbins. Red dashed lines mark $\pm 1\sigma$ bounds, where $\sigma$ is estimated using median absolute deviation (MAD). Typical residual scatter is $\lesssim 1$ km/s across all statistics—smaller than observational uncertainties.

Figure \ref{fig:NN_vb_75} demonstrates physical consistency by showing how $\langle v_b \rangle$ and $\sigma_{v_b}$ vary with the mass of the LMC ($M_{\mathrm{LMC}}$) and the Milky Way ($M_{\mathrm{MW}}$) for stars in the 60–90 kpc radial range. As $M_{\mathrm{LMC}}$ increases from $0.7$ to $2.1 \times 10^{11}{\rm M}_\odot$, $\langle v_b \rangle$ nearly doubles (from $\sim$9 to $\sim$17 km/s), while $\sigma_{v_b}$ increases only 3\% (from $\sim$68 to $\sim$70 km/s). In contrast, when $M_{\mathrm{MW}}$ increases from $0.5$ to $1.0 \times 10^{12}{\rm M}_\odot$, $\langle v_b \rangle$ decreases by about 7\% (from $\sim$14 to $\sim$13 km/s), whereas $\sigma_{v_b}$ increases more significantly by approximately 24\% (from $\sim$62.5 to $\sim$77.5 km/s). This confirms that first moments respond strongly to LMC perturbations while second moments primarily trace MW halo structure.

For parameter inference, we define the likelihood:
\begin{equation}
    \mathcal{L}(\boldsymbol{\theta}) = \exp\left( -\frac{1}{2} \sum_k \frac{(y_{\text{obs},k} - y_{\text{NN},k}(\boldsymbol{\theta}))^2}{\sigma_k^2} \right)
\label{eq:likelihood}
\end{equation}
where $y_{\text{obs},k}$ is the observed statistic, $y_{\text{NN},k}(\boldsymbol{\theta})$ is the neural network prediction, and $\sigma_k$ accounts for observational uncertainty (emulation error is negligible).

We model our mock observations on typical halo tracers such as RR Lyrae stars, given their abundance and brightness ($M_\mathrm{G} \simeq +0.6$), which enable measurements to $\sim$120 kpc. Following the density profile from \citet{2018ApJ...859...31H} and Section \ref{sec:stellar_tagging}, we assume a total of 7,000 stars within 20–131 kpc. Of these, approximately 4,000 stars fall within our analysis range of 30–120 kpc: $\sim$3,000 in 30–60 kpc, $\sim$700 in 60–90 kpc, and $\sim$300 in 90–120 kpc. These numbers are broadly consistent with the observed RR Lyrae population in recent catalogs \citep[e.g.,][]{2023ApJ...944...88L}, although we note that the observed sample becomes increasingly incomplete beyond 100 kpc. 

Observational uncertainties in summary statistics are computed analytically, as detailed in Appendix~\ref{appendix:sampling_error}, by propagating individual measurement errors while accounting for finite sample sizes in each radial bin. For tracers like RR Lyrae, distances achieve $\sim$10\% relative uncertainty \citep[e.g.,][]{10.1093/mnras/sty2241,2022MNRAS.513..788G,2023ApJ...944...88L}. Radial velocities range from a few km/s (DESI) to tens of km/s (SDSS) for faint stars; we assume 20 km/s typical uncertainty. Tangential velocity precision depends on distance and proper motion:
\begin{equation}
\frac{\epsilon_v}{\mathrm{~km} \mathrm{~s}^{-1}}=\frac{\epsilon_D}{D} \frac{v}{\mathrm{~km} \mathrm{~s}^{-1}}+4.74 \frac{\epsilon_\mu}{\mathrm{mas~yr}^{-1}} \frac{D}{\mathrm{kpc}}
\end{equation}
with $\epsilon_D/D = 10\%$ and heliocentric velocities $\sim$200-300 km/s. PyGaia \citep{2021A&A...649A...3R} estimates proper motion uncertainties: $\sim$0.1-1.0 mas yr$^{-1}$ (DR3) improving to $\sim$0.02-0.2 mas yr$^{-1}$ (DR5) for typical halo tracers at 30–120 kpc.

Figure \ref{fig:Gaia_uncertainty} (left) shows tangential velocity uncertainties versus distance for Gaia DR3, DR4 (2× improvement), and DR5 (5× improvement). Beyond 60 kpc, DR3 uncertainties exceed several hundred km/s. The right panel shows the signal-to-noise ratio (SNR) of tangential velocity dispersion, comparing the stellar halo's physical velocity dispersion ($\sim$60-70 km/s from the fiducial model) to the statistical sampling uncertainty (scaling as $\sim 1/\sqrt{2N}$ where N decreases with distance; see Appendix~\ref{appendix:sampling_error}). Each curve incorporates the fixed measurement precision of its respective data release. Effective parameter constraints require SNR $\gtrsim$ 5–10. DR3's SNR falls below 5 beyond $\sim$60 kpc, while DR4 and DR5 extend this to $\sim$70 and $\sim$90 kpc. Where tangential velocities become noise-dominated, radial velocities provide crucial constraints.

To demonstrate that our kinematic summary statistics can successfully recover MW-LMC parameters, we first present a proof-of-concept analysis using a specific observational configuration. We adopt Gaia DR3-like tangential velocity uncertainties and assume 20 km/s radial velocity precision as a representative case. While Section \ref{sec:forecast} will systematically explore how different observational scenarios affect parameter constraints using Fisher matrix forecasts, here we validate that our MCMC-based inference framework can reliably recover the true parameters from mock observations.

Figure~\ref{fig:posterior_corner} presents posterior distributions for $M_{\mathrm{MW}}$, $M_{\mathrm{LMC}}$, $c$, and $q$ using nested sampling (\texttt{dynesty}; \citealt{2020MNRAS.493.3132S}). Mock observations include mean velocities and dispersions in radial (separated by hemisphere), latitudinal, and longitudinal directions for three distance bins.

To isolate the smooth halo, we exclude: circular regions of $20^{\circ}$ and $10^{\circ}$ around the LMC and SMC \citep{2001AJ....122.1807V, 2000A&A...358L...9C}; stars within $|b_{\text{Sgr}}| < 15^{\circ}$ of the Sagittarius plane \citep{2014MNRAS.437..116B, 2021Natur.592..534C}; and stars with $|b| < 10^{\circ}$ to minimize extinction. We account for hemispheric asymmetry with $N_{b>0^{\circ}}:N_{b<0^{\circ}}$ ratios of 1.1, 1.3, and 1.3 for the three distance bins.

The marginalized posteriors yield: $M_{\mathrm{MW}} = 0.64^{+0.12}_{-0.08} \times 10^{12}{\rm M}_\odot$, $M_{\mathrm{LMC}} = 17.00^{+3.04}_{-2.57} \times 10^{10}{\rm M}_\odot$, $c = 10.91^{+2.38}_{-2.48}$, and $q = 0.97^{+0.06}_{-0.05}$. All parameters recover the true values within 1–2$\sigma$, demonstrating the method's potential for constraining the MW–LMC system.

\subsection{Forecasting Parameter Constraints}
\label{sec:forecast}

Having validated our neural network framework's ability to recover MW–LMC parameters from mock observations, we now use Fisher matrix analysis to forecast the method's constraining power across various observational scenarios. This approach efficiently explores how different data quality improvements—from Gaia DR3 to DR5 astrometry, radial velocity inclusion, sample size, and distance precision—affect parameter constraints.

\subsubsection{Fisher Matrix Methodology and Validation}
\label{sec:fisher_method}

\begin{figure}
	\includegraphics[width=1\columnwidth]{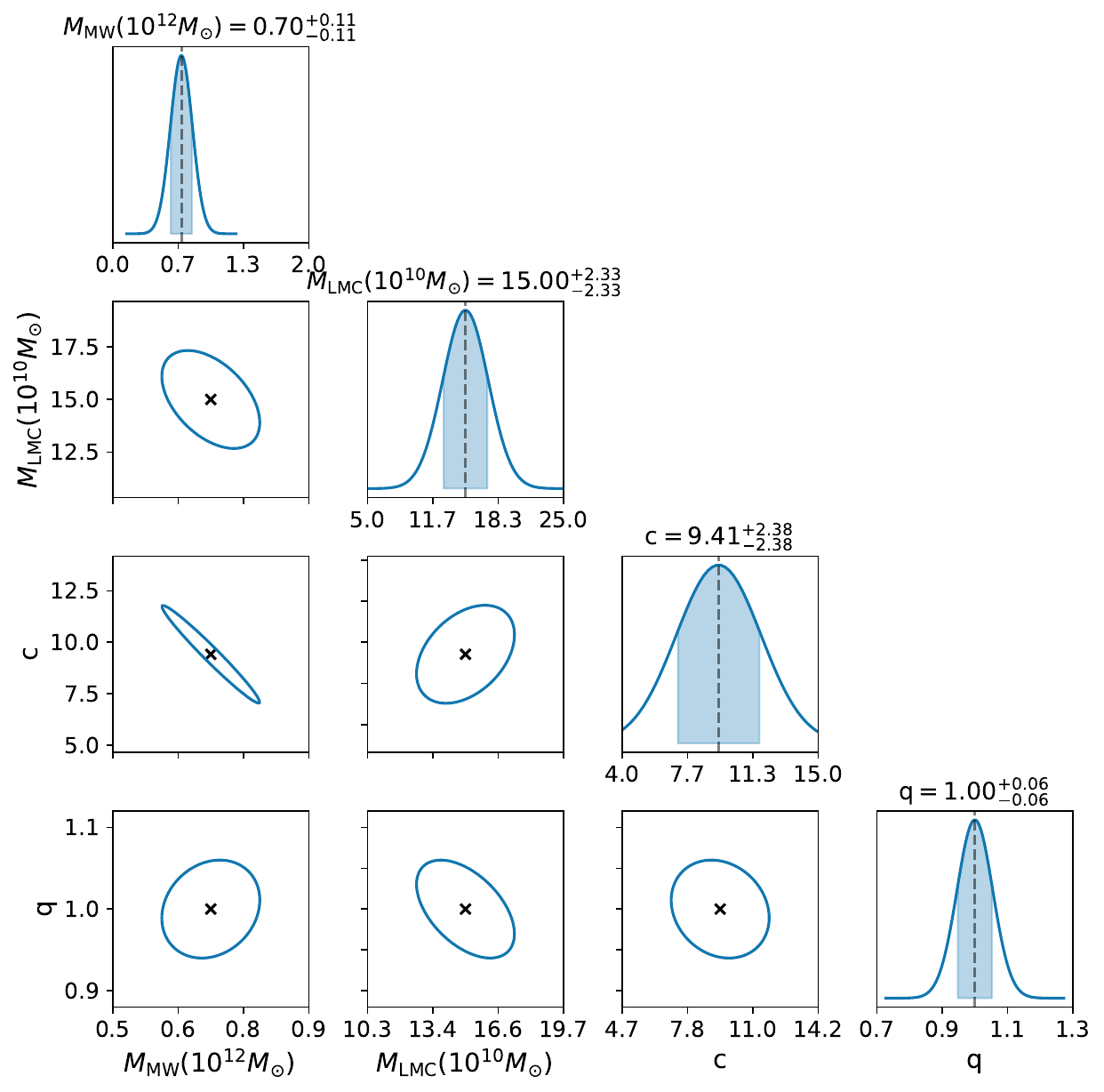}
    \caption{Predicted covariance ellipses for $M_{\mathrm{MW}}$, $M_{\mathrm{LMC}}$, $c$, and $q$ computed at the fiducial parameter combination using the Fisher matrix. We incorporate observational uncertainties that include both measurement errors (20 km/s for stellar radial velocities and Gaia DR3–like uncertainties for tangential velocities) and finite sampling noise from the assumed stellar population. The diagonal panels show the 1D marginalized posterior distributions, with shaded regions indicating the 1$\sigma$ credible intervals: $0.11 \times 10^{12} {\rm M}_\odot$ for $M_{\mathrm{MW}}$, $2.33 \times 10^{10} {\rm M}_\odot$ for $M_{\mathrm{LMC}}$, 2.38 for $c$, and 0.06 for $q$. The off-diagonal panels display the 1$\sigma$ confidence ellipses for the 2D joint posterior distributions. The close agreement between these Fisher ellipses and the posterior contours from Section~\ref{sec:posterior_inference} confirms the validity of the Fisher matrix approximation, supporting the use of the Fisher matrix for forecasting.}
    \label{fig:posterior_vs_fisher}
\end{figure}

\begin{figure}
	\includegraphics[width=1\columnwidth]{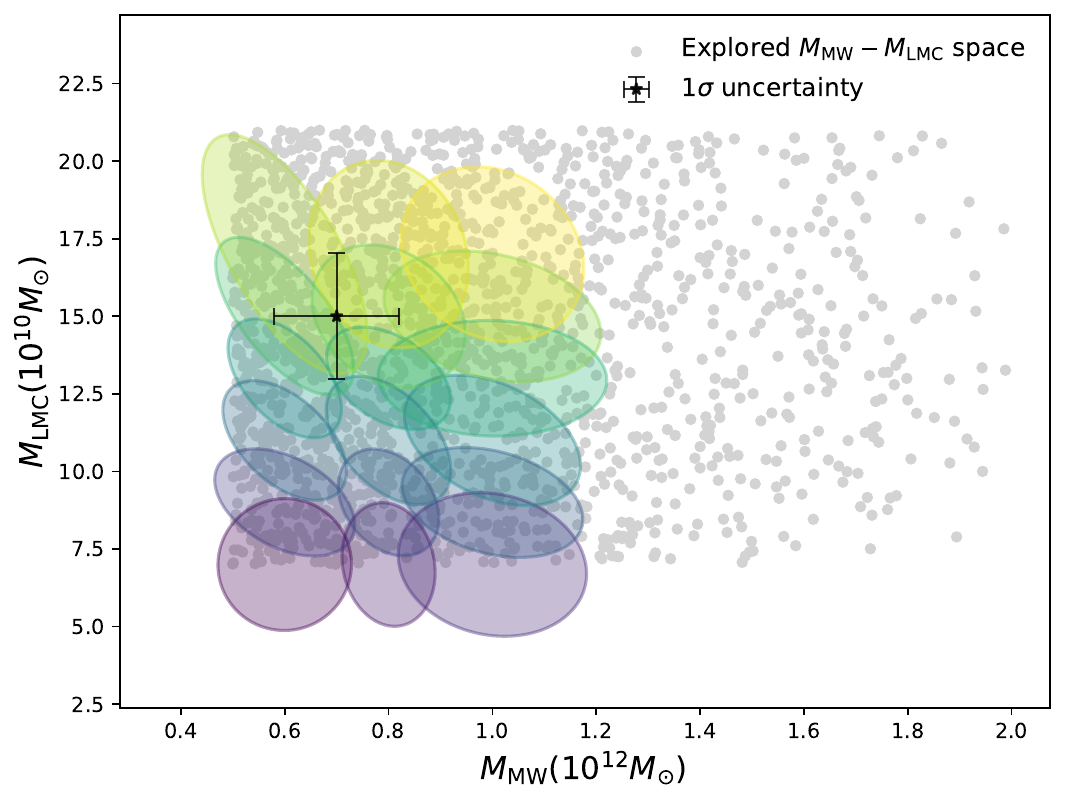}
    \caption{Forecast 1$\sigma$ covariance ellipses computed using the Fisher matrix across a grid of parameter combinations in the $(M_{\mathrm{MW}}, M_{\mathrm{LMC}})$ plane, assuming the same observational uncertainties used in Figure~\ref{fig:posterior_vs_fisher} (20 km/s radial velocity precision and Gaia DR3-like tangential velocity uncertainties, including finite sampling noise). The background grey dots are the $\sim$2,000 parameter combinations from our simulation suite. Each ellipse represents the predicted uncertainty at a specific parameter combination, with colors varied decoratively to improve the visibility of overlapping regions. The grid spans $M_{\mathrm{MW}} = 0.5,0.7,0.9\times 10^{12} {\rm M}_{\odot}$ and $M_{\mathrm{LMC}} = 7-17 \times 10^{10} {\rm M}_{\odot}$, with other parameters fixed at fiducial values ($c = 9.415$ and $q = 1.0$). While uncertainties remain broadly consistent across the grid, an increase of up to 1.6 times in the $\sigma_{M_{\mathrm{LMC}}}$ is observed near the edges of the parameter space. The symbol with error bars indicates the fiducial values of $M_{\mathrm{MW}}$ and $M_{\mathrm{LMC}}$, along with their 1$\sigma$ forecast uncertainties from the marginalized posterior distribution.}
    \label{fig:Fisher_ellipses}
\end{figure}

The posterior distributions in Section~\ref{sec:posterior_inference} demonstrated parameter recovery for one specific noise realization. To assess the general constraining power across different observational configurations and parameter values, we employ the Fisher matrix formalism, which provides the expected parameter precision without requiring full posterior sampling for each scenario.

The Fisher matrix elements are (see Appendix~\ref{appendix:Fisher matrix method} for derivation):
\begin{equation}
F_{ij} = \sum_k \frac{1}{\sigma_k^2} \frac{\partial y_{\mathrm{NN},k}}{\partial \theta_i} \frac{\partial y_{\mathrm{NN},k}}{\partial \theta_j},
\end{equation}
where $\sigma_k$ represents the observational uncertainty (including measurement errors and finite sampling noise), and derivatives $\partial y_{\mathrm{NN},k} / \partial \theta_i$ are computed numerically using central finite differences on the smooth neural network emulation. This approach enables rapid exploration of parameter space regions beyond our simulation grid.

Figure~\ref{fig:posterior_vs_fisher} validates the Fisher matrix by comparing its predictions with the posterior distributions from Section~\ref{sec:posterior_inference}. The diagonal panels show 1D marginal distributions with 1$\sigma$ intervals: $0.11 \times 10^{12} {\rm M}_\odot$ for $M_{\mathrm{MW}}$, $2.33 \times 10^{10} {\rm M}_\odot$ for $M_{\mathrm{LMC}}$, 2.38 for $c$, and 0.06 for $q$. Off-diagonal panels display 2D confidence ellipses. The agreement between Fisher forecasts and posterior contours confirms the validity of the Fisher matrix approximation.

Figure~\ref{fig:Fisher_ellipses} extends this validation to a parameter grid in the $(M_{\mathrm{MW}}, M_{\mathrm{LMC}})$ plane. Since each Fisher matrix is constructed from local derivatives of the observables, changing the reference point modifies these derivatives. If the observables become less sensitive to a given parameter at a new location, the corresponding diagonal element $F_{ii}$ of the Fisher matrix decreases, leading to a larger forecast uncertainty $\sigma_{\theta_i} \propto F_{ii}^{-1/2}$. Conversely, stronger local sensitivities tighten the constraint. These shifts therefore reflect variations in local information content.

The $1\sigma$ uncertainties remain nearly consistent (typically $\simeq0.10\times10^{12}M_\odot$ for $M_{\mathrm{MW}}$ and $\simeq(2$–$3)\times10^{10}M_\odot$ for $M_{\mathrm{LMC}}$) across most of the grid. A moderate growth of up to $1.6\times$ in $\sigma_{M_{\mathrm{LMC}}}$ appears at the high-mass edge, where the response of the observables to $M_{\mathrm{LMC}}$ flattens and the Fisher information declines (see the top-left panel of Figure \ref{fig:NN_vb_75}). We show results only up to $M_{\mathrm{MW}} = 1.0 \times 10^{12} M_\odot$, as higher MW masses—when combined with the fiducial concentration ($c = 9.415$) and flattening ($q = 1.0$)—place the LMC on a second-passage rather than a first-infall orbit. 

\subsubsection{Forecasting Precision Under Different Observational Scenarios}

\begin{figure}
	\includegraphics[width=1\columnwidth]{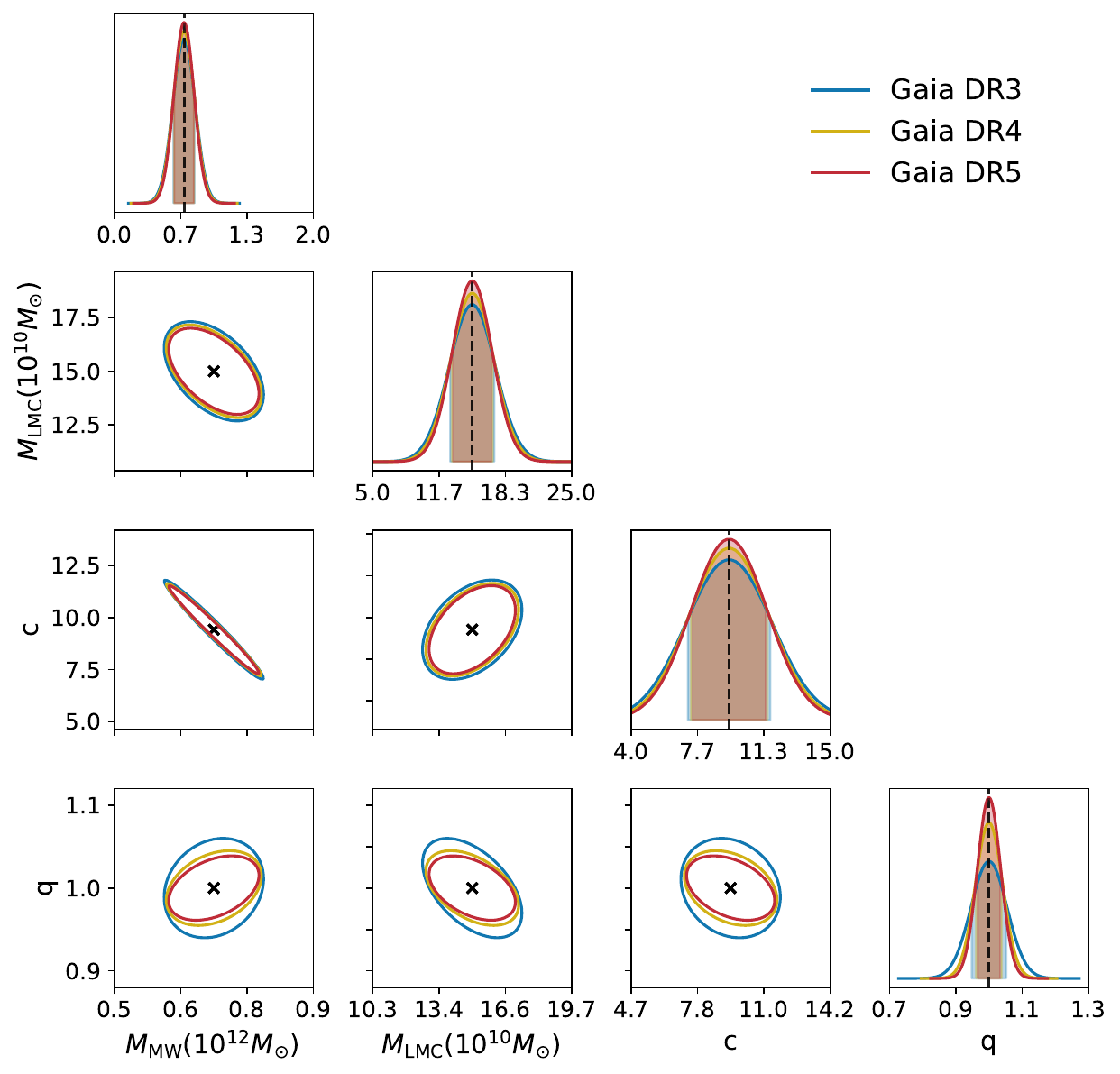}
    \caption{Forecast 1D marginalized distributions (diagonal panels) and 1$\sigma$ confidence ellipses (off-diagonal panels) for $M_{\mathrm{MW}}$, $M_{\mathrm{LMC}}$, $c$, and $q$, computed using the full set of kinematic summary statistics under Gaia DR3 (blue), DR4 (yellow), and DR5 (red) tangential velocity precisions. The Fisher matrix is evaluated at the fiducial parameter combination. Improvements in Gaia astrometry from DR3 to DR4 (DR5) reduce forecast uncertainties by approximately 6\% (12\%) for $M_{\mathrm{MW}}$, 7\% (13\%) for $M_{\mathrm{LMC}}$, 9\% (14\%) for $c$ and 27\% (38\%) for $q$. These results show that improvements in Gaia astrometric precision lead to only moderate enhancements in parameter constraints, particularly when the radial velocity uncertainty is already small (e.g., 20 km/s).}
    \label{fig:Fisher_precision}
\end{figure}

\begin{figure}
	\includegraphics[width=1\columnwidth]{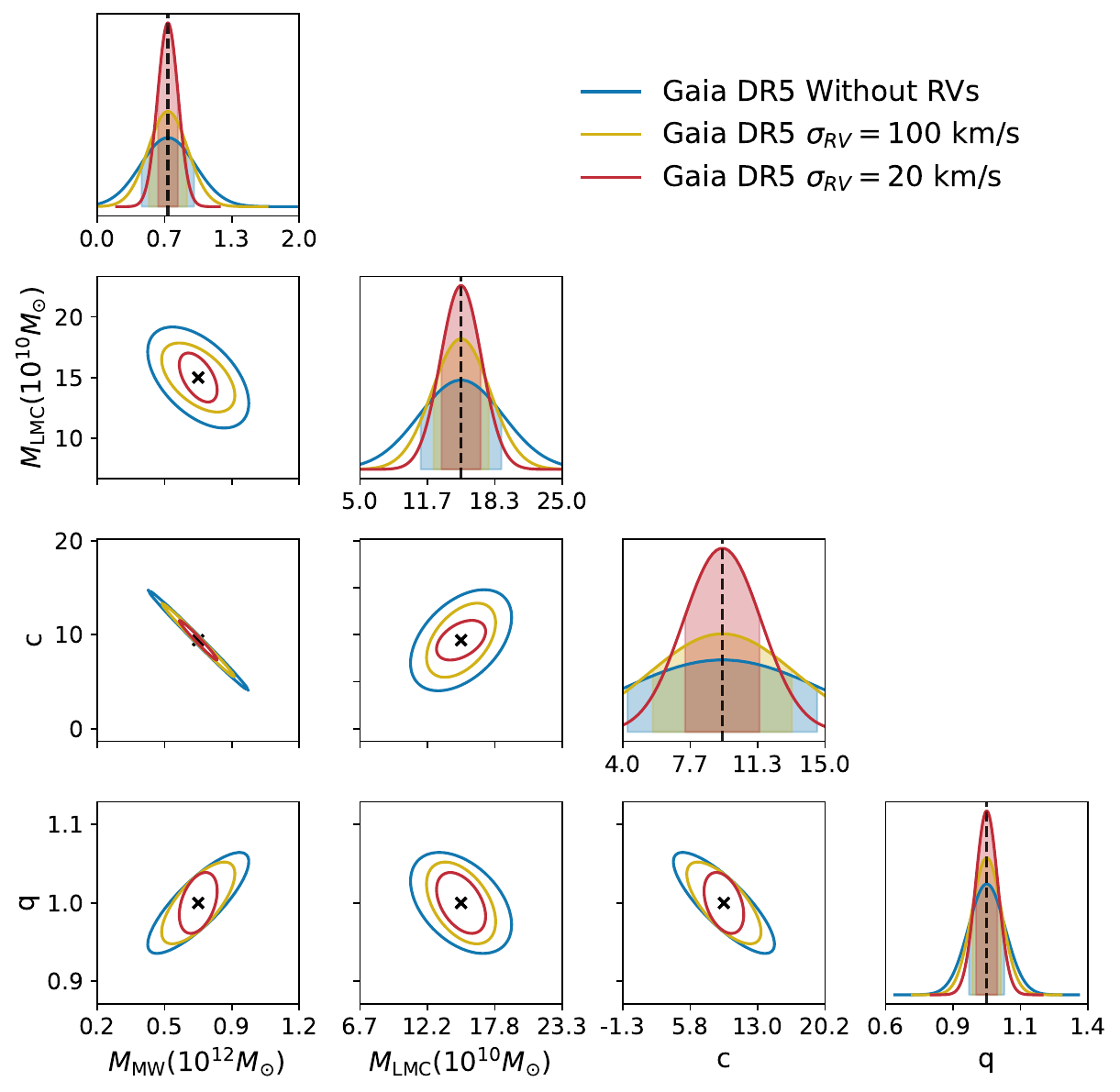}
    \caption{ Forecast parameter constraints for $M_{\mathrm{MW}}$, $M_{\mathrm{LMC}}$, $c$, and $q$ under Gaia DR5 astrometric precision, comparing three scenarios: full summary statistics with radial velocity (RV) uncertainty of 20 km/s (red contours), with RV uncertainty of 100 km/s (yellow contours), and excluding RV measurements entirely (blue contours). Including RVs significantly improves forecast precision. Compared to the no-RV case, incorporating RVs with $\sigma_{\mathrm{RV}} = 100$ km/s improves constraints by approximately 27\% for $M_{\mathrm{MW}}$, 33\% for $M_{\mathrm{LMC}}$, and 26\% for $c$. With improved RV precision of 20 km/s, the corresponding improvements increase to about 60\%, 50\%, and 60\%, respectively. In contrast, the constraint on $q$ remains largely unaffected across these scenarios. These results underscore the critical role of RV measurements in constraining the masses and concentration of the MW and LMC.}
    \label{fig:Fisher_RVs}
\end{figure}

\begin{figure}
	\includegraphics[width=1\columnwidth]{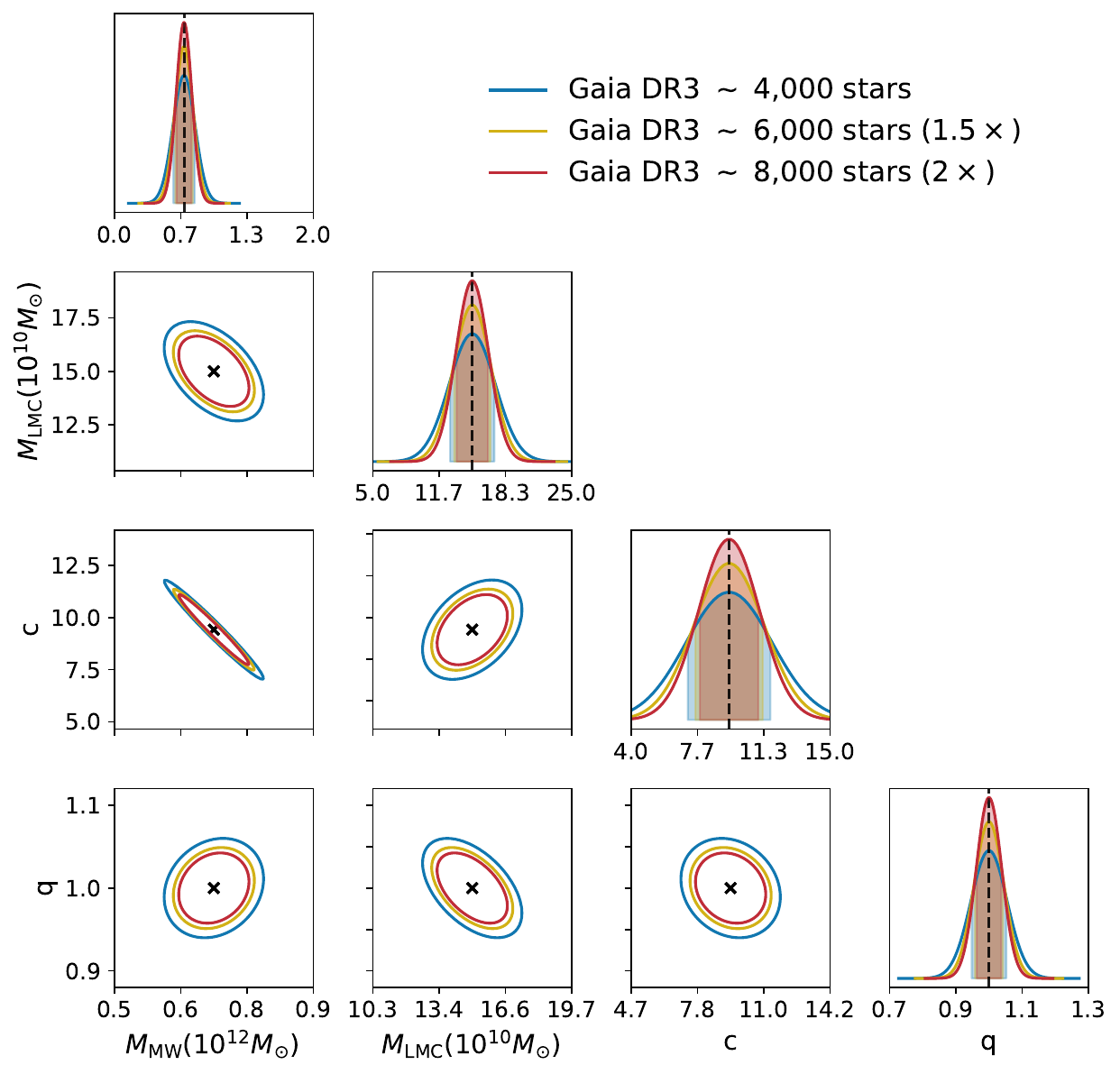}
    \caption{Forecast constraints for $M_{\mathrm{MW}}$, $M_{\mathrm{LMC}}$, $c$, and $q$ as a function of sample size, assuming Gaia DR3 astrometric precision, 10\% distance uncertainties, and 20 km/s radial velocity precision. The baseline in our study adopts $\sim$4,000 RR Lyrae stars within the 30–120 kpc range (blue), with distance distribution consistent with the known RR Lyrae population. We compare this to scenarios with increased sample sizes of 6,000 (yellow) and 8,000 (red) stars, drawn from the same underlying density profile. Parameter uncertainties scale approximately as $1/\sqrt{N}$: the 6,000-star sample yields $\sim$20\% tighter constraints, and the 8,000-star sample yields $\sim$30\% improvements relative to the baseline. The gains are most pronounced beyond 60 kpc, where current RR Lyrae samples are especially limited. These results underscore the value of expanding outer halo tracer catalogs for dynamical inference.}
    \label{fig:Fisher_samplesize}
\end{figure}

\begin{figure}
	\includegraphics[width=1\columnwidth]{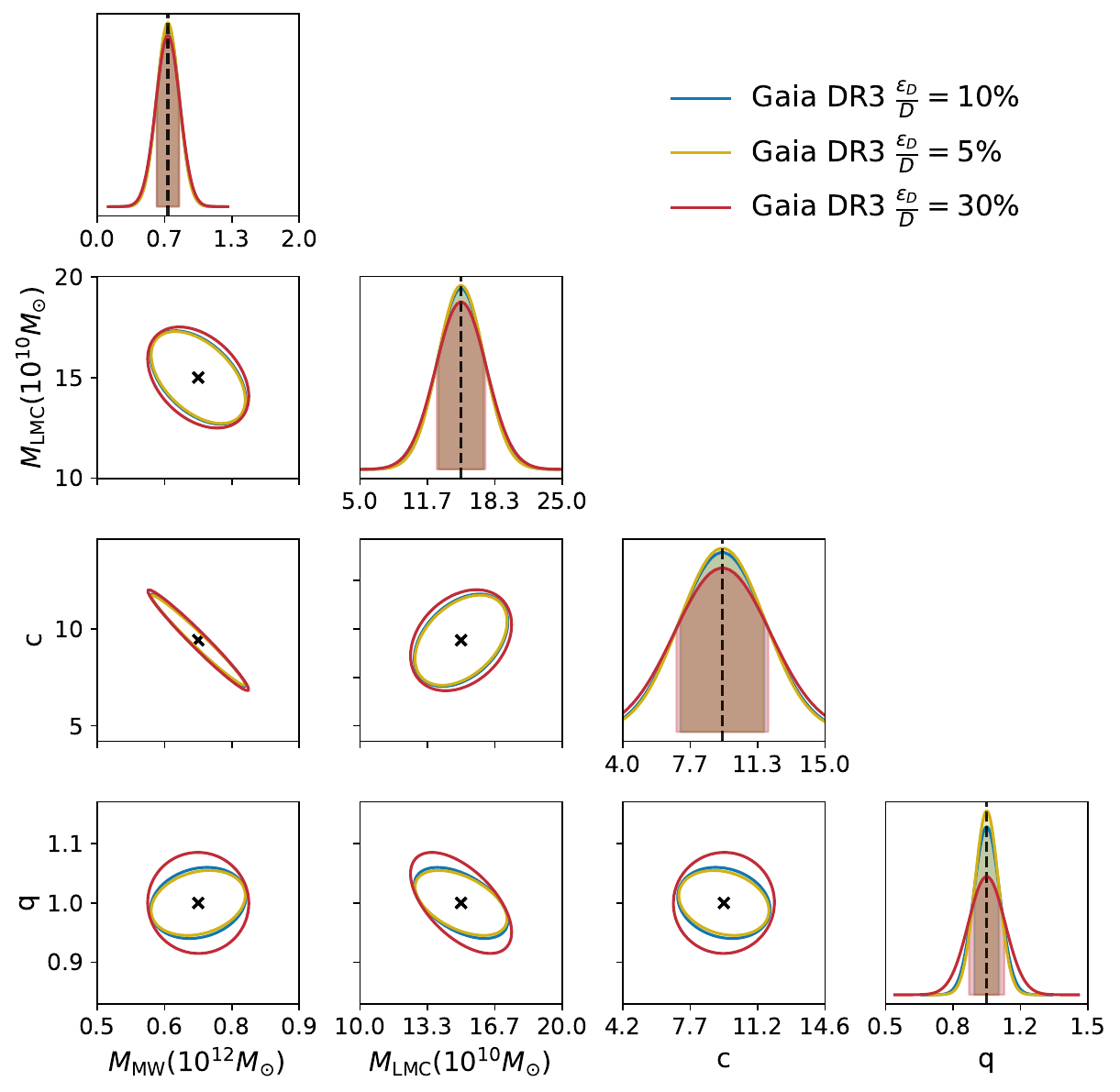}
    \caption{Impact of photometric distance precision on parameter constraints for $M_{\mathrm{MW}}$, $M_{\mathrm{LMC}}$, $c$, and $q$, assuming Gaia DR3 astrometric precision, 20 km/s radial velocity uncertainty, and $\sim$4,000 RR Lyrae stars. We compare three scenarios: 5\% distance uncertainty (yellow), 10\% (blue; typical for RR Lyrae), and 30\% (red; typical for RGB stars). Constraints on $M_{\mathrm{MW}}$, $M_{\mathrm{LMC}}$, and $c$ degrade by only $\sim$10\% when increasing the distance uncertainty from 10\% to 30\%, while improving from 10\% to 5\% yields a marginal $\sim$2–3\% gain. The weak dependence of parameter constraints on distance uncertainties demonstrates the robustness of our method even for halo tracers with less precise distances.}
    \label{fig:Fisher_distance}
\end{figure}

The Fisher matrix formalism allows us to readily evaluate and visualize how observational improvements affect parameter constraints, which we explore below while avoiding the computational burden of posterior sampling. Below, we present both absolute uncertainties and their fractional precision relative to the fiducial values of $M_{\mathrm{MW}} = 1.0 \times 10^{12} M_\odot$, $M_{\mathrm{LMC}} = 1.5 \times 10^{11} M_\odot$, $c = 9.415$, and $q = 1.0$.

\textbf{Astrometric precision from Gaia DR3 to DR5.} Figure~\ref{fig:Fisher_precision} shows forecast constraints using the full kinematic summary statistics (mean velocities and dispersions in radial, latitudinal, and longitudinal directions for three distance bins) under Gaia DR3 (blue), DR4 (yellow), and DR5 (red) precisions, with fixed 20 km/s radial velocity uncertainty, 10\% distance errors and a sample of $\sim$4,000 RR Lyrae stars between 30-120 kpc. For Gaia DR3, the $1\sigma$ uncertainties are $0.11\times10^{12}M_\odot$, $2.33\times10^{10}M_\odot$, 2.38, and 0.06 for $M_{\mathrm{MW}}$, $M_{\mathrm{LMC}}$, $c$, and $q$, corresponding to fractional precisions of 11\%, 16\%, 25\%, and 6\% respectively. Moving to DR4 improves constraints to 11\%, 14\%, 24\% and 5\%, while DR5 to 10\%, 13\%, 22\% and 4\%, reflecting modest gains of 6–14\% in mass and concentration, and up to 38\% in flattening. These modest gains reflect the dominance of the 20 km/s radial velocity precision. Without radial velocities—common in the outer halo where spectroscopy is limited—only DR5-level precision enables converged posteriors.

\textbf{The role of radial velocity measurements.} Figure~\ref{fig:Fisher_RVs} compares constraints under Gaia DR5 astrometry with fixed 10\% distance errors, a sample of $\sim$4,000 stars, while varying radial velocity scenarios: 20 km/s uncertainty (red), 100 km/s uncertainty (yellow), and no radial velocities (blue). We focus on DR5 because larger tangential velocity uncertainties in DR3/DR4 prevent posterior convergence without radial velocities, violating the Fisher matrix's approximation.

Including radial velocities dramatically improves constraints. Without RVs, the $1\sigma$ uncertainties are $0.27\times10^{12}M_\odot$, $4.16\times10^{10}M_\odot$, 5.38 and 0.06 for $M_{\mathrm{MW}}$, $M_{\mathrm{LMC}}$, $c$ and $q$, corresponding to fractional precisions of 27\%, 28\%, 57\% and 6\%. Adding RVs with even 100 km/s precision (comparable to Euclid expectations; \citealt{2025A&A...697A...1E}) reduces these to $0.20\times10^{12}M_\odot$, $2.84\times10^{10}M_\odot$, 3.95 and 0.05 (20\%, 19\%, 42\% and 5\%). With 20 km/s precision, the constraints further tighten to $0.10\times10^{12}M_\odot$, $2.02\times10^{10}M_\odot$, 2.11 and 0.04 (10\%, 13\%, 22\% and 4\%). The constraint on halo flattening $q$ remains relatively unaffected by the inclusion or quality of RV measurements.

\textbf{Sample size effects.} Figure~\ref{fig:Fisher_samplesize} presents the impact of sample size on parameter constraints. Our baseline assumes $\sim$4,000 RR Lyrae stars within 30-120 kpc, comparable to current datasets \citep[e.g.,][]{2023ApJ...944...88L}. Since summary statistic uncertainties scale as $1/\sqrt{N}$, larger samples directly improve constraints. Increasing to $\sim$6,000 stars (1.5×) improves precision by $\sim$20\% for all parameters, reducing uncertainties to $0.09\times10^{12}M_\odot$, $1.90\times10^{10}M_\odot$, 1.94 and 0.05 (9\%, 13\%, 21\% and 5\%). Doubling to $\sim$8,000 stars yields $\sim$30\% improvement, with uncertainties of $0.08\times10^{12}M_\odot$, $1.64\times10^{10}M_\odot$, 1.68 and 0.04 (8\%, 11\%, 18\% and 4\%). These gains are especially significant beyond 60 kpc where current samples are sparse ($\sim$1,000 stars) and tangential velocity uncertainties are large. 

\textbf{Distance precision.} Figure~\ref{fig:Fisher_distance} illustrates how distance uncertainties affect the forecast precision. Current photometric distances for RR Lyrae achieve $\sim$10\% uncertainty. Improving to 5\% yields only 2–3\% better constraints on $M_\mathrm{MW}$, $M_\mathrm{LMC}$, and $c$. Even degrading to 30\% precision (typical for red giants) worsens constraints by only $\sim$10\%, demonstrating weak sensitivity to distance uncertainties.

\section{Discussion}
\label{sec:discussion}

In this study, we investigated how all-sky kinematic summary statistics of MW halo stars—specifically the first moments (mean velocities) and second moments (velocity dispersions)—encode information about the MW–LMC interaction. Using a suite of high-resolution simulations ($\sim10^{7}$ particles), we trained a neural network to emulate these summary statistics as functions of four key parameters: $M_{\mathrm{MW}}$, $M_{\mathrm{LMC}}$, $c$, and $q$. We then examined how observational uncertainties in stellar radial velocities, distances, and proper motions—reflecting the precision expected from Gaia DR3 through DR5—as well as finite sampling noise affect constraints on these parameters.

Our results reveal that first and second moments carry complementary information about the MW-LMC system. The mean velocities respond strongly to perturbations induced by the LMC, making them particularly sensitive to the LMC mass. In contrast, velocity dispersions primarily reflect the intrinsic structure of the MW halo and thus provide better constraints on MW properties. We further find that including radial velocity measurements and increasing stellar sample sizes improve parameter constraints, whereas improvements in distance precision yield more moderate gains.

\subsection{Implications for current and future surveys}
\label{sec:survey_implications}

Our Fisher matrix forecasts provide guidance for optimizing observational strategies to constrain MW-LMC properties. The sensitivity analysis reveals a clear hierarchy of observational improvements, with important implications for ongoing and planned surveys.

The most gains come from including radial velocity measurements. Even with modest precision of 100 km/s—comparable to what Euclid will achieve for faint halo tracers \citep{2025A&A...697A...1E}—parameter constraints improve by 30\% compared to using Gaia astrometry alone. In this case, the forecasted $1\sigma$ uncertainties reach $0.20 \times 10^{12} M_\odot$, $2.84 \times 10^{10} M_\odot$, 3.95, and 0.05 for $M_{\mathrm{MW}},\ M_{\mathrm{LMC}},\ c$, and $q$, corresponding to fractional uncertainties of 20\%, 19\%, 42\%, and 5\% respectively, relative to the fiducial values adopted in this study. This suggests that Euclid's all-sky coverage and ability to reach the outer halo will provide valuable complementary data to Gaia, despite its relatively poor velocity precision. For brighter targets where spectroscopic follow-up achieves 20 km/s precision, constraints improve by 50–60\%, reaching $0.10 \times 10^{12} M_\odot$, $2.02 \times 10^{10} M_\odot$, 2.11, and 0.04 (10\%, 13\%, 22\%, and 4\%), demonstrating the high value of dedicated spectroscopic campaigns. Other upcoming space missions like CSST \citep{2021RAA....21...92S} and Roman \citep{2022ApJ...928....1W} (with its grism spectroscopy mode) may provide similar capabilities, though their final spectroscopic specifications for stellar science remain to be determined. 

Ground-based spectroscopic surveys are particularly well-suited for this science. DESI \citep{2019BAAS...51g..57L}, with its ability to reach faint magnitudes and obtain spectra for millions of stars, can probe the outer halo where LMC-induced perturbations are strongest. The survey's typical radial velocity precision of 10-20 km/s for fainter halo tracers provides the measurements our analysis shows are crucial for breaking parameter degeneracies. Similarly, LAMOST's \citep{2012RAA....12..723Z} large field of view and systematic coverage of the northern sky has already produced radial velocities for hundreds of thousands of halo stars. 

The H3 survey \citep{2019ApJ...883..107C} provides another valuable all-sky dataset with high-resolution spectroscopy of halo stars, while targeted surveys like S5 \citep{2019MNRAS.490.3508L} have focused on specific streams and substructures. The recently planned VIA survey \citep{Via-project} is obtaining high-precision radial velocities and detailed abundances for stars across the Galaxy. The upcoming 4MOST \citep{de20164most} and WEAVE \citep{2024MNRAS.530.2688J} surveys will further expand this dataset, with 4MOST's and WEAVE's wide-area coverage providing complementary approaches to mapping the MW-LMC interaction.

Sample size emerges as the second most important factor. Doubling the number of halo tracers from current levels ($\sim$4,000) to $\sim$8,000 improves constraints by 30\%, reaching $0.08\times10^{12}M_\odot$, $1.64\times10^{10}M_\odot$, 1.68 and 0.04 (8\%, 11\%, 18\% and 4\%). This is readily achievable with existing and upcoming surveys. For instance, combining samples from different tracer populations (RR Lyrae, BHB stars, red giants) or extending magnitude limits can easily double or triple available samples. The gains are particularly pronounced beyond 60 kpc where current samples are sparse.

Interestingly, our analysis shows that distance precision improvements yield minimal gains—only 2-3\% improvement when reducing uncertainties from 10\% to 5\%. This suggests that current photometric distance estimates are already sufficient for this application. Even using tracers with 30\% distance uncertainties (like faint red giants) degrades constraints by only 10\%. This opens opportunities to use less precise but more numerous tracers without compromising results.

The modest improvements from Gaia DR3 to DR5 astrometry (10-15\% for most parameters) indicate that we are approaching diminishing returns from proper motion improvements alone, at least for samples with good radial velocities. However, for studies relying solely on Gaia data, the DR5 improvements remain crucial—only at DR5 precision can tangential velocities alone provide meaningful constraints in the outer halo.

These findings suggest an optimal observational strategy: prioritize radial velocity measurements over astrometric precision improvements, expand samples by including multiple tracer populations even if they have larger distance uncertainties, and focus spectroscopic efforts on the outer halo (beyond 60 kpc) where perturbations are strongest. The combination of space-based wide-area surveys like Euclid, CSST, and Roman with ground-based spectroscopic surveys (DESI, LAMOST, 4MOST, WEAVE, H3, VIA) will provide the most advances in constraining the MW-LMC interaction.

\subsection{Interpretation of first and second moment sensitivities}
\label{sec:interpretation}

Our simulations show that the mean velocity and velocity dispersion of halo stars respond quite differently to perturbations induced by the LMC. To understand this behavior, we develop a simplified analytical model that captures the essential physics of the MW's reflex motion.

In this model, we consider that the inner regions of the MW are accelerating toward a location along the LMC's past orbit. This acceleration creates a coherent motion pattern in the halo that manifests as a dipole velocity field: from the perspective of an observer in the inner MW, distant halo stars appear to be moving upward in the Galactocentric frame as the MW itself moves downward toward the LMC's trajectory.

The bulk velocity vector of this dipole motion can be expressed as:
\begin{equation}
\mathbf{v}_{\text {dipole }}=v_{\text {dipole }}(\cos i, 0, \sin i)
\end{equation}
where $i$ is the inclination (or apex latitude) of the dipole vector relative to the Galactic disk plane. For an observer at the center of the Galaxy, the latitudinal velocity component in a given sky direction $(\theta,\phi)$ becomes:
\begin{equation}
v_b(\theta, \varphi)=v_{\text {dipole }}[-\sin \theta \cos \varphi \cos i+\cos \theta \sin i]
\end{equation}

By integrating this expression over all sky directions and assuming an isotropic distribution of halo stars, we can derive the mean latitudinal velocity:
\begin{equation}
\left\langle v_b\right\rangle=\frac{1}{4 \pi} \int_0^{2 \pi} \int_{-\pi / 2}^{\pi / 2} v_b(\theta, \varphi) \cos \theta \mathrm{d} \theta \mathrm{~d} \varphi=\frac{\pi}{4} v_{\text {dipole }} \sin i
\end{equation}

This result reveals that the mean latitudinal velocity scales directly with the dipole velocity amplitude $v_{\mathrm{dipole}}$. Since $v_{\mathrm{dipole}}$ itself is proportional to the mass of the LMC (as shown in Figure 6 of \citet{2021NatAs...5..251P}), this explains why mean velocities are particularly sensitive to $M_{\mathrm{LMC}}$.

The situation is quite different for velocity dispersions. The variance induced solely by the dipole motion is:
\begin{equation}
\sigma_{b, \text { dipole }}^2=\left\langle v_b^2\right\rangle-\left\langle v_b\right\rangle^2=v_{\text {dipole }}^2\left[\frac{1}{6} \cos ^2 i+\left(\frac{2}{3}-\frac{\pi^2}{16}\right) \sin ^2 i\right]
\end{equation}

However, this dipole-induced variance must be combined with the intrinsic velocity dispersion of the halo. Since the intrinsic stellar velocities are uncorrelated with the coherent dipole motion, the dispersions add in quadrature:
\begin{equation}
\sigma_{b, \text { total }}=\sqrt{\sigma_0^2+\sigma_{b, \text { dipole }}^2}
\end{equation}
where $\sigma_{0}$ is the intrinsic one-dimensional velocity dispersion in the latitudinal direction.

To illustrate the relative importance of these effects, consider typical values from observations: an intrinsic halo velocity dispersion $\sigma_{0}$=75 km/s, an apex latitude $i=-45^{\circ}$, and a dipole velocity amplitude $v_{\mathrm{dipole}}=30$ km/s \citep{2021NatAs...5..251P,2024arXiv240601676C,2024arXiv241009149B}. Under these conditions, the mean latitudinal velocity induced by the MW's reflex motion is $\langle v_{b}\rangle \approx 17$ km/s—a clear and measurable signal. In contrast, the total velocity dispersion increases only by 1 km/s, barely distinguishable from the intrinsic dispersion.

This example demonstrates why first and second moments provide complementary constraints. The dipole motion generates a substantial shift in mean velocities that scales with $M_{\mathrm{LMC}}$, while its contribution to velocity dispersions remains negligible unless $v_{\mathrm{dipole}}$ becomes comparable to $\sigma_{0}$. Consequently, velocity dispersions remain dominated by the intrinsic kinematics of the MW halo and thus provide better sensitivity to MW structural parameters like $M_{\mathrm{MW}}$, $c$, and $q$.

\subsection{Impact of LMC trajectory uncertainties on stellar summary statistics}
\label{sec:LMC_trajectory_statistics}

An important systematic uncertainty in our analysis arises from reconstructing the LMC's past trajectory after it entered the MW's virial radius. In Section~\ref{sec:orbital reconstruction}, we developed a neural network-based forward model that learns how the LMC's phase-space coordinates evolve over time for different MW-LMC parameter combinations. This allows us to infer the initial conditions 2 Gyr ago that lead to the observed present-day position and velocity. Although our forward-integrated trajectories generally reproduce the observed coordinates within 2$\sigma$, residual discrepancies remain that could affect our kinematic predictions.

To quantify this effect, we selected five representative simulations from our parameter grid that showed the largest deviations between the neural network predictions and the observed LMC phase-space coordinates. For each case, we applied the Gauss–Newton iteration method \citep[following][]{2020MNRAS.497.4162V,2024MNRAS.527..437V,2024MNRAS.534.2694S} to refine the initial conditions and minimize the present-day mismatch. We then compared the velocity summary statistics calculated using the original neural network-derived trajectories versus the refined trajectories (see Appendix~\ref{appendix:trajectory_uncertainties} for detailed comparisons).

For most statistics, the variations are reassuringly small ($\lesssim 1$ km/s). However, we find that the mean radial velocity of stars in the southern hemisphere ($b<0^{\circ}$) can show more substantial changes. In the most extreme case (Table~\ref{tab:LMC_orbit_variation}, row 4), the refined trajectory differs from the neural network prediction by $\Delta R_{\mathrm{init}}\approx 47$ kpc and $\Delta V_{\mathrm{init}}\approx 27$ km/s at the initial epoch. This produces a change of $\Delta \langle v_{r,b<0^{\circ}} \rangle \sim 8$ km/s for stars at 60-90 kpc—a shift that exceeds the expected measurement uncertainties at these distances, assuming a per-star RV precision of 20 km/s.

Even simulations with smaller orbital corrections can produce noticeable effects. For instance, trajectory differences of only $\Delta R_{\mathrm{init}}\approx 4$ kpc and $\Delta V_{\mathrm{init}}\approx 4$ km/s can still yield $\Delta \langle v_{r,b<0^{\circ}} \rangle \sim 5$ km/s at 60–90 kpc (Table~\ref{tab:LMC_orbit_variation}, row 2).

These results reveal that the relationship between orbital uncertainties and velocity statistics is non-linear. The impact depends on how specific trajectory perturbations alter the MW's reflex motion and the subsequent dynamical response of the halo. While the mean radial velocity in the southern hemisphere shows potential sensitivity to trajectory uncertainties, it represents only one of approximately 20 summary statistics used in our analysis. The vast majority of these statistics—including all velocity dispersions and most mean velocities—vary by less than 1 km/s under trajectory refinement, well below typical measurement uncertainties. 

This indicates that our overall parameter constraints remain reliable despite the sensitivity of this particular statistic. Nevertheless, future work should explore methods to either down-weight or exclude $\langle v_{r,b<0^{\circ}} \rangle$ in regions where trajectory uncertainties dominate, and continue improving LMC trajectory modeling to minimize this source of systematic uncertainty.

\subsection{Caveats and future directions}
\label{sec:limitations}

\begin{figure}
	\includegraphics[width=\columnwidth]{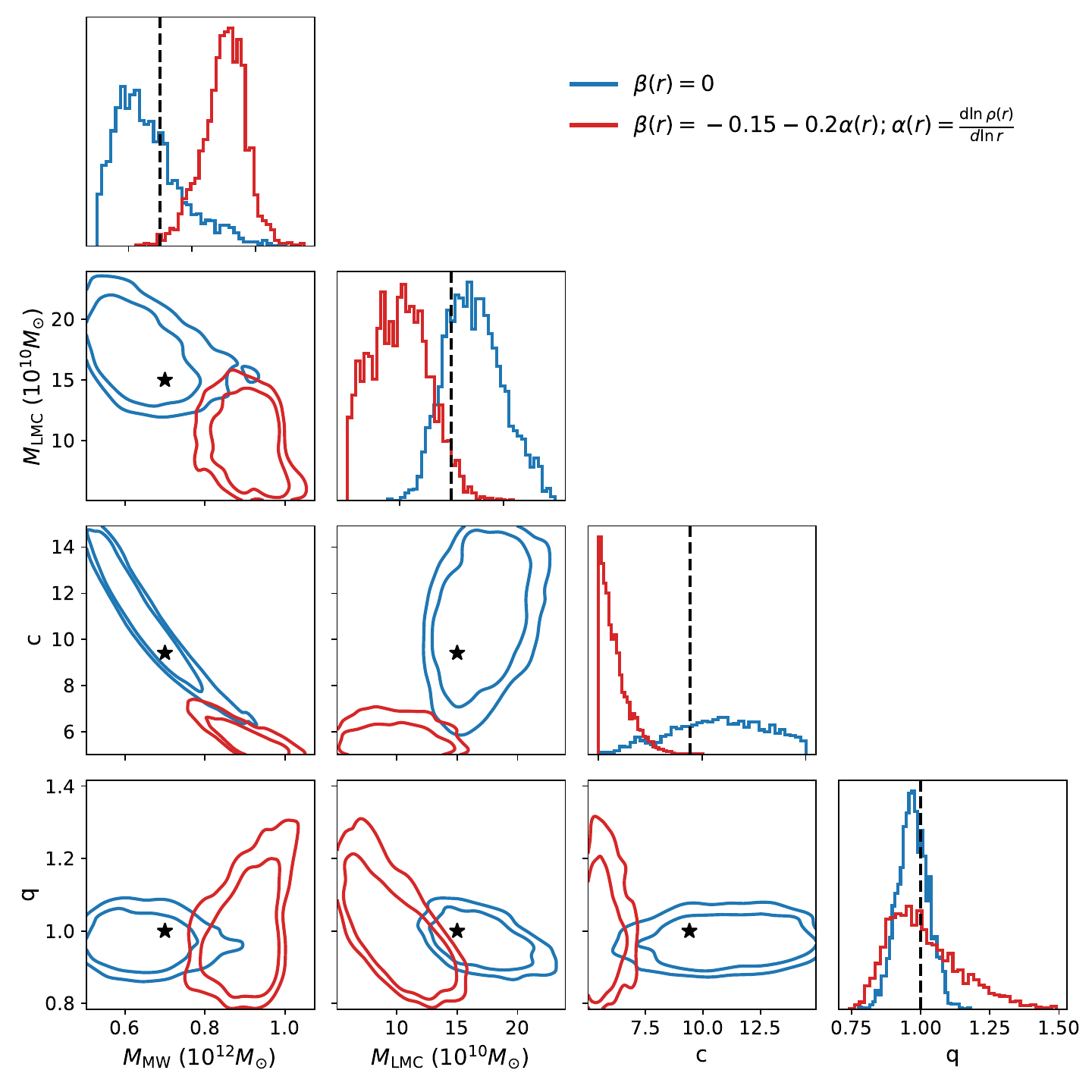}
    \caption{Posterior distributions of the MW–LMC model parameters under two different assumptions for the velocity anisotropy profile of halo stars. Contours represent the 1$\sigma$ and 1.5$\sigma$ confidence levels, assuming Gaussian equivalents. The blue contours correspond to the correct model with constant anisotropy $\beta(r) = 0$, while the red contours correspond to an incorrect model with radially varying anisotropy $\beta(r) = -0.15 - 0.2\alpha(r)$, where $\alpha(r) = \mathrm{d} \ln \rho(r)/\mathrm{d} \ln r$. Black dashed lines mark the true parameter values in the 1D marginalized distributions (diagonal panels), and black stars indicate the true values in the 2D contours (off-diagonal panels). The comparison reveals large biases in the recovered parameters when anisotropy is incorrectly modeled: $M_{\mathrm{MW}}$ is overestimated by approximately 40\%, $M_{\mathrm{LMC}}$ is underestimated by a similar fraction, and $c$ is biased toward lower values, falling outside the original 1$\sigma$ credible intervals.}
    \label{fig:posterior_beta_compare}
\end{figure}

\begin{figure}
	\includegraphics[width=\columnwidth]{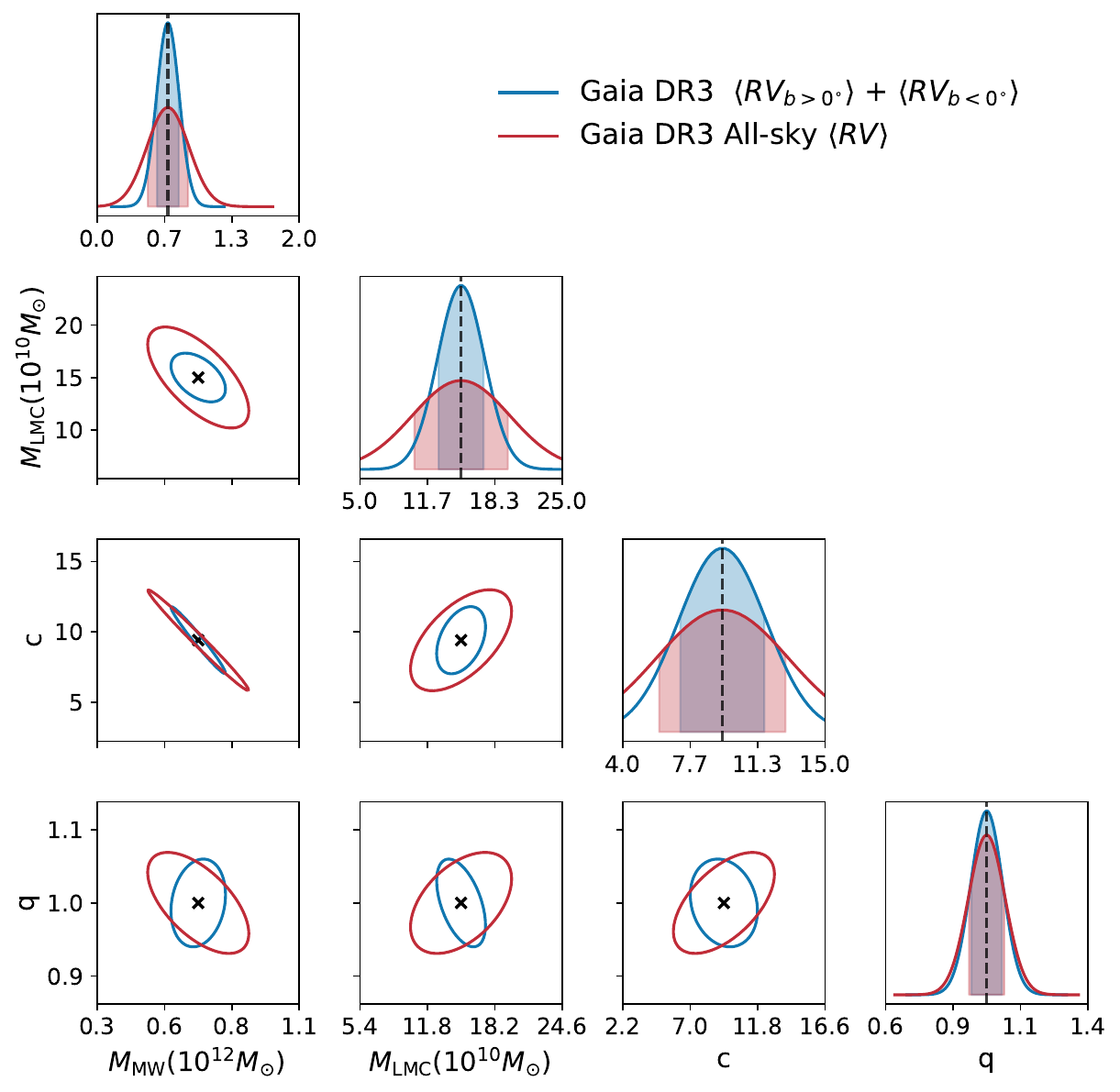}
    	\caption{Impact of spatial binning on forecast parameter constraints. This figure compares Fisher matrix forecast ellipses for $M_{\mathrm{MW}}$, $M_{\mathrm{LMC}}$, $c$ and $q$ under Gaia DR3-level proper motion precision. The red contours represent forecasts obtained by separately modeling the mean radial velocities in the northern and southern Galactic hemispheres to account for the dipole-like structure induced by the MW's reflex motion. The blue contours show forecasts when using only the all-sky averaged radial velocity, neglecting the hemispheric asymmetry. Neglecting the dipole pattern degrades the parameter constraints, nearly doubling the forecast uncertainties for $M_{\mathrm{MW}}$, $M_{\mathrm{LMC}}$, and $c$, highlighting the additional constraining power contained in the spatial structure of the kinematic perturbations.}
    \label{fig:Fisher_RVs_all_sky}
\end{figure}

While our simulation-based approach demonstrates the potential for constraining MW-LMC properties through stellar kinematics, several simplifying assumptions may affect the precision and applicability of our results. Here we discuss these limitations and outline directions for future work.

\subsubsection{Simplified N-body models of the Milky Way} 

Our MW models use a simplified version of \texttt{MWPotential2014} \citep{2015ApJS..216...29B}, varying only dark matter halo properties while fixing disk and bulge components. This enables efficient equilibrium initial conditions via \textsc{galic} but omits important structural features. Unlike the McMillan potential \citep{2017MNRAS.465...76M}, we lack a thick disk and gas disk that contribute to vertical forces at intermediate heights ($|z| \sim 1$–2 kpc). Similarly, we omit the central nucleus component included in Price-Whelan models \citep{2017JOSS....2..388P}. While these omissions minimally affect our region of interest (beyond 30 kpc), they limit applicability to inner-halo dynamics.

Our constant-shape, axisymmetric NFW halo aligned with the disk oversimplifies the likely MW halo structure. \citet{2021MNRAS.501.2279V} demonstrate that a double-power-law profile with radially varying axis ratios better reproduces features like the Sgr stream. The real MW halo may also be tilted and triaxial due to the Gaia–Sausage–Enceladus merger \citep{2022AJ....164..249H,2022ApJ...934...14H}. However, constructing such complex models in dynamical equilibrium for live N-body simulations remains technically challenging.

The velocity anisotropy profile $\beta(r)$ represents another critical simplification. While our main grid assumes isotropy ($\beta=0$), we also ran 1,000 simulations with radially varying anisotropy. Though both models produce similar halo responses to LMC perturbations, the inferred parameters—particularly from velocity dispersions—depend sensitively on the assumed $\beta(r)$.

To quantify this sensitivity, we performed a revealing test: we generated mock observations from our fiducial simulation with $\beta(r) = 0$, then applied our inference pipeline using the emulation model trained on simulations with radially varying $\beta(r)$. Figure~\ref{fig:posterior_beta_compare} shows the dramatic consequences of this mismatch. The incorrect anisotropy assumption biases $M_{\mathrm{MW}}$ high by $\sim$40\%, $M_{\mathrm{LMC}}$ low by $\sim$40\%, and pushes $c$ well outside the original 1$\sigma$ intervals. These biases arise because radial and tangential dispersions depend on orbital anisotropy—when $\beta(r)$ is misspecified, the model compensates by adjusting other parameters to match the observed dispersions.

This result underscores that fixing the anisotropy profile can yield misleading parameter constraints. Future studies should either marginalize over flexible $\beta(r)$ forms or incorporate observational priors on anisotropy.

\subsubsection{Absence of the SMC} 

We model the LMC as a single spherical halo, ignoring the Small Magellanic Cloud despite its $\sim$10\% mass ratio to the LMC. While the SMC has minimal direct impact on MW halo kinematics, it can perturb the LMC's trajectory, reducing its orbital period and apocenter distance \citep{2023Galax..11...59V}. Modeling the three-body MW-LMC-SMC interaction in live N-body simulations remains challenging, particularly for determining initial conditions that match current observations. Our simplification effectively includes the SMC mass within the LMC.

Our neural network approach, while efficient for two-body MW-LMC interactions, faces limitations when extended to three-body systems. The higher-dimensional parameter space and stricter observational constraints make it difficult to find initial coordinates that reproduce present-day positions within required uncertainties. Future work could explore Physics-Informed Neural Networks (PINNs) \citep{2018arXiv180607366C,2020arXiv200504849H} which embed differential equations directly into the network architecture. Rather than learning point mappings, PINNs learn the governing dynamics, treating network layers as time evolution steps. For MW-LMC-SMC systems, this would transform orbit reconstruction from simulation emulation to physics-constrained inference, providing a fully differentiable model that adjusts initial coordinates until trajectories match observed error ellipsoids for both clouds.

\subsubsection{Beyond all-sky averages} 

Our framework uses first and second moments—mean velocities and dispersions—which are straightforward to compute with analytical error estimates. However, Figures~\ref{fig:fiducial_mean_map} and~\ref{fig:fiducial_std_map} reveal finer spatial structures along the LMC's path not captured by all-sky statistics. These coherent perturbations suggest that fitting the full 6D phase-space distribution could tighten constraints.

Figure~\ref{fig:Fisher_RVs_all_sky} illustrates this potential. Comparing global mean radial velocity (blue) with hemisphere-separated means accounting for the dipole structure (red), we find that neglecting the dipole nearly doubles forecast uncertainties for $M_{\mathrm{MW}}$, $M_{\mathrm{LMC}}$, and $c$ under Gaia DR3 precision.

While higher-moment statistics offer greater constraining power, they require better observational precision to overcome increased sensitivity to measurement errors. Simulation-based inference (SBI) \citep[e.g.,][]{2019MNRAS.488.4440A,2019ApJ...876....3L,doi:10.1073/pnas.1912789117,2023ApJ...942...26G,2025JCAP...01..021L,2025arXiv250710667B} provides a promising path forward, using neural density estimation to approximate probability distributions for halo density and kinematic features, without resorting to summary statistics as explored in this study. Future work will explore applying SBI to the MW-LMC-SMC system.

\section{Conclusion}
\label{sec:conclusions}

LMC-induced dynamical effects in the MW halo provide fundamental constraints on the MW-LMC system, including mass distributions and orbital characteristics. Understanding these properties is essential for interpreting the MW's assembly history, calibrating models of satellite galaxy interactions, and testing predictions of galaxy formation in a cosmological context. In this first paper of a series, we have developed a comprehensive suite of 2,848 high-resolution ($10^7$ particles) N-body simulations systematically exploring variations in MW mass profile and LMC infall mass. We model the mean velocity and velocity dispersion of halo stars across different Galactic radii. We then assess how these kinematic summary statistics can constrain MW-LMC properties given various observational scenarios, providing forecasts for current and future capabilities. Our main conclusions are:

\begin{itemize}
    \item Mean velocities and velocity dispersions probe complementary physics. Mean velocities capture LMC-induced bulk motions, providing strong sensitivity to $M_{\mathrm{LMC}}$. Velocity dispersions reflect the MW halo's intrinsic potential, constraining $M_{\mathrm{MW}}$, $c$, and $q$. Joint analysis breaks parameter degeneracies that would persist if either observable were used in isolation.
    
    \item Using full kinematic statistics (radial, latitudinal, and longitudinal components across 30–60, 60–90, and 90–120 kpc) and assuming 10\% distance precision, 20 km/s radial velocity accuracy, and a sample of $\sim$4,000 RR Lyrae stars, the forecast $1\sigma$ uncertainties are $0.11\times10^{12}M_\odot$, $2.33\times10^{10}M_\odot$, 2.38, and 0.06 for $M_{\mathrm{MW}}$, $M_{\mathrm{LMC}}$, $c$, and $q$, corresponding to fractional precisions of 11\%, 16\%, 25\%, and 6\%. Improvements in Gaia proper motion precision from DR3 to DR5 yield modest forecast gains: 6–12\% for $M_{\mathrm{MW}}$, 7–13\% for $M_{\mathrm{LMC}}$, 9–14\% for $c$, and 27–38\% for $q$. These moderate improvements suggest that astrometric precision is approaching diminishing returns when combined with radial velocity data.
    
    \item Radial velocities are crucial despite observational challenges. Under Gaia DR5 precision, adding 100 km/s RVs improves constraints by $\sim$30\% for $M_{\mathrm{MW}}$, $M_{\mathrm{LMC}}$, and $c$ compared to astrometry alone, reaching $1\sigma$ uncertainties of $0.20 \times 10^{12} M_\odot$, $2.84 \times 10^{10} M_\odot$, 3.95 and 0.05 (20\%, 19\%, 42\% and 5\%). With 20 km/s RV precision, constraints improve by $\sim$50–60\%, reaching $0.10 \times 10^{12} M_\odot$, $2.02 \times 10^{10} M_\odot$, 2.11 and 0.04 (10\%, 13\%, 22\% and 4\%). This dramatic enhancement underscores the value of spectroscopic surveys for MW halo science.
    
    \item Sample size strongly affects constraints through reduced sampling noise. Assuming Gaia DR3 astrometry, 10\% distance precision, and 20 km/s radial velocity uncertainty, increasing the number of halo tracers from $\sim$4,000 to $\sim$6,000 improves forecast precision by $\sim$20\%; doubling to $\sim$8,000 yields a $\sim$30\% improvement, with $1\sigma$ uncertainties reaching $0.08\times10^{12}M_\odot$, $1.64\times10^{10}M_\odot$, 1.68, and 0.04 (8\%, 11\%, 18\%, and 4\%) for $M_{\mathrm{MW}}$, $M_{\mathrm{LMC}}$, $c$, and $q$ respectively. Gains are largest beyond 60 kpc where tangential velocity uncertainties dominate. Distance precision has minimal impact: improving from 10\% to 5\% yields only 2–3\% gains, while degrading to 30\% costs only $\sim$10\%. This suggests that numerous tracers with moderate distance precision are preferable.
    
    \item Systematic biases arise from two main sources. First, mismodeling the velocity anisotropy profile $\beta(r)=-0.15-0.2 \alpha(r) ; \alpha(r)=\frac{\mathrm{d} \ln \rho(r)}{d \ln r}$ causes large parameter biases (e.g., 40\% errors in masses) since anisotropy controls the radial-tangential dispersion partitioning. Second, mean radial velocities in the southern hemisphere ($b<0^\circ$) are sensitive to LMC trajectory reconstruction uncertainties. Future work should incorporate spatially localized features and marginalize over anisotropy profiles to reduce model dependence.
\end{itemize}

Our forecasts demonstrate that halo stellar kinematics can precisely constrain the MW-LMC system, particularly the virial masses and MW density profile. The simulation suite and inference framework developed here provide the foundation for interpreting observations. This methodology paper establishes the theoretical groundwork and quantifies the expected constraining power under various observational scenarios. 

In the next paper of this series, we will apply this framework to existing spectroscopic datasets, including LAMOST, DESI DR1, and the forthcoming Euclid DR1, to derive the first constraints on MW-LMC parameters using our comprehensive simulation grid. The combination of these surveys with Gaia proper motions and additional spectroscopic programs will provide complete 6D phase-space information from 30 to beyond 100 kpc. Integrating these datasets with the simulation-based inference framework presented here will enable increasingly precise constraints on MW-LMC parameters and reveal the full spatial structure of LMC-induced perturbations throughout the Milky Way halo. Such measurements will ultimately constrain the MW's total mass, the LMC's orbital history, and the nature of dynamical friction in realistic galactic environments—key ingredients for understanding galaxy evolution in the Local Group and beyond.

\section*{Acknowledgements}

We are grateful to the Scientific and Local Organizing Committees of the XMC II: Clouds over Yellowstone workshop for providing an excellent platform to present our results. The workshop fostered valuable discussions and offered the opportunity to engage with many experts in the field, from whom we received insightful feedback and constructive advice that helped shape the direction of future works. Y.S.T is supported by the National
Science Foundation under Grant No. AST-2406729. X.-X.X. acknowledges the support from the National Key Research and Development Program of China No. 2024YFA1611902, National Natural Science Foundation of China (NSFC) No. 12588202, CAS Project for Young Scientists in Basic Research grant No. YSBR-062, the Strategic Priority Research Program of Chinese Academy of Sciences grant No. XDB1160102 and grant No. CMS-CSST-2025-A11. We further acknowledge the high performance computing resources provided by the Australian National Computational Infrastructure (grants y89) through the National and ANU Computational Merit Allocation Schemes. 

This work has made use of software developed by the Gaia Project Scientist Support Team and the Gaia Data Processing and Analysis Consortium (DPAC). Funding for the DPAC as been provided by national institutions, in particular the institutions participating in the Gaia Multilateral Agreement. 

\textit{Software}: IPython \citep{4160251}, matplotlib \citep{4160265}, numpy \citep{harris2020array}, scipy \citep{2020SciPy-NMeth}, Astropy \citep{astropy:2013,astropy:2018}, Gadget-4 \citep{2021MNRAS.506.2871S}, GALIC \citep{2014MNRAS.444...62Y}, galpy \citep{2015ApJS..216...29B}, h5py \citep{collette_python_hdf5_2014}, gala \citep{2017JOSS....2..388P}, dynesty \citep{2020MNRAS.493.3132S}, PyGaia \citep{2021A&A...649A...3R}.

\section*{Data Availability}

The N-body simulation suite developed in this work, HaloDance, will be made publicly available via GitHub at \href{https://github.com/Yanjun-Sheng/HaloDance}{\texttt{github.com/Yanjun-Sheng/HaloDance}}. These include 101 snapshots spanning the past 2 Gyr, and can be applied for stellar stream modeling, halo kinematics, and forward modeling of Gaia observations.



\bibliographystyle{mnras}
\bibliography{example} 





\appendix

\section{Fisher matrix forecast methodology}
\label{appendix:Fisher matrix method}

The Fisher matrix approach efficiently evaluates how well future observations can constrain MW-LMC parameters. We define the parameter vector as:
\begin{equation}
\boldsymbol{\theta}=\left(M_{\mathrm{MW}}, M_{\mathrm{LMC}}, c, q\right)
\end{equation}

Let $y_{k}$ denote the $N$ kinematic summary statistics (mean velocities and velocity dispersions). Since these statistics show negligible correlation, we treat them as independent. The log-likelihood for parameters $\boldsymbol{\theta}$ is:
\begin{equation}
\ln \mathcal{L}(\boldsymbol{\theta})=-\frac{1}{2} \sum_k\left[\ln \left(2 \pi \sigma_k^2\right)+\Delta y_k^2 / \sigma_k^2\right],\quad \Delta y_k \equiv y_{\text{obs},k}-y_{\text{NN},k}(\boldsymbol{\theta})
\end{equation}
where $y_{\text{obs },k}$ are observed values, $y_{\text{NN }, k}$ are neural network predictions, and $\sigma_k^2$ represents the total observational uncertainty including measurement errors and finite sampling noise.

The Fisher matrix equals the expectation of the negative Hessian:
\begin{equation}
F_{i j}=-\left\langle\frac{\partial^2 \mathcal{L}}{\partial \theta_i \partial \theta_j}\right\rangle=\sum_k \frac{1}{\sigma_k^2} \frac{\partial y_{\mathrm{NN}, k}}{\partial \theta_i} \frac{\partial y_{\mathrm{NN}, k}}{\partial \theta_j}
\label{eq:Fisher_derivation}
\end{equation}

This represents the Laplace approximation to the posterior. The covariance matrix $\mathbf{C}=\mathbf{F}^{-1}$ provides forecast uncertainties: $\sigma\left(\theta_i\right)=\sqrt{C_{i i}}$ for individual parameters, with off-diagonal terms indicating correlations.

The derivatives $\partial y_{\mathrm{NN}, k} / \partial \theta_i$ are computed numerically using central finite differences on the smooth neural network emulation. This approach enables rapid exploration of parameter space regions beyond our discrete simulation grid, making it valuable for assessing how observational improvements affect parameter constraints.

\section{Propagation of Individual Velocity Errors to Sample-Level Summary Statistics}
\label{appendix:sampling_error}

Summary statistics inherit uncertainty from both individual measurement errors and finite sample size. We derive how these propagate to the mean velocity and velocity dispersion, accounting for both sources of uncertainty.

\textbf{Mean velocity.} Each star's observed velocity is:
\begin{equation}
y_i=\mu+\varepsilon_i+e_i,
\end{equation}
where $\mu$ is the population mean, $\varepsilon_i \sim \mathcal{N}(0, \sigma_{\text {int }}^2)$ represents intrinsic scatter, and $e_i \sim \mathcal{N}(0, \sigma^2)$ is measurement error.

The sample mean $\widehat{\mu} = \frac{1}{N}\sum y_i$ has variance:
\begin{equation}
\operatorname{Var}[\widehat{\mu}]=\frac{1}{N^2} \sum_{i=1}^N \operatorname{Var}\left(y_i\right)=\frac{\sigma_{\text {int }}^2+\sigma^2}{N}
\end{equation}

Therefore, the total uncertainty in the mean velocity is:
\begin{equation}
\delta \mu_{\mathrm{tot}}=\sqrt{\frac{\sigma_{\mathrm{int}}^2+\sigma^2}{N}}
\end{equation} 

This expression shows how both intrinsic dispersion and measurement error contribute to the uncertainty, scaled by the sample size.

\textbf{Velocity dispersion.} We estimate the intrinsic dispersion as $\widehat{\sigma}_{\text {int }}^2=s^2-\sigma^2$, where $s^2$ is the sample variance:
\begin{equation}
s^2=\frac{1}{N-1} \sum_{i=1}^N\left(y_i-\bar{y}\right)^2
\end{equation}

From Cochran's theorem \citep{1934PCPS...30..178C}, $(N-1) s^2/\sigma_{\text {tot }}^2 \sim \chi_{N-1}^2$, where $\sigma_{\text {tot}}^2 = \sigma_{\text{int}}^2+\sigma^2$. This gives:
\begin{equation}
\operatorname{Var}\left[s^2\right]=\frac{2 \sigma_{\text{tot }}^4}{N-1}=\frac{2\left(\sigma_{\text {int }}^2+\sigma^2\right)^2}{N-1}
\end{equation}

Since $\widehat{\sigma}_{\text {int }}^2=s^2-\sigma^2$ and $\sigma$ is assumed constant:
\begin{equation}
\operatorname{Var}\left[\widehat{\sigma}_{\text {int }}^2\right]=\frac{2\left(\sigma_{\text {int }}^2+\sigma^2\right)^2}{N-1}
\end{equation}

Applying the delta method with $g(x)=\sqrt{x}$:
\begin{equation}
\operatorname{Var}\left[\widehat{\sigma}_{\mathrm{int}}\right] \approx\left(\frac{1}{2 \sigma_{\mathrm{int}}}\right)^2 \operatorname{Var}\left[\widehat{\sigma}_{\mathrm{int}}^2\right]=\frac{\left(\sigma_{\mathrm{int}}^2+\sigma^2\right)^2}{2(N-1) \sigma_{\mathrm{int}}^2}
\end{equation}

The uncertainty on intrinsic dispersion is:
\begin{equation}
\delta \sigma_{\mathrm{int}}=\frac{\sigma_{\mathrm{int}}^2+\sigma^2}{\sigma_{\mathrm{int}}} \sqrt{\frac{1}{2(N-1)}}
\end{equation}

These expressions account for both measurement errors and finite sampling in realistic observational scenarios. For practical calculations, we fix $\sigma_{\text{int}}$ to values from our fiducial MW model, which provides the expected intrinsic velocity dispersion at each Galactocentric radius.

\section{LMC trajectory reconstruction uncertainties}
\label{appendix:trajectory_uncertainties}

As discussed in Section~\ref{sec:LMC_trajectory_statistics}, our neural network approach for reconstructing the LMC's initial conditions from present-day observations can introduce systematic uncertainties in the predicted stellar kinematics. While the neural network generally reproduces observed LMC phase-space coordinates within 2$\sigma$, residual discrepancies can propagate into the velocity summary statistics used for parameter inference.

To quantify these effects, we selected five representative MW-LMC parameter combinations from our simulation grid that exhibited the largest deviations between neural network predictions and observations. For each case, we applied Gauss-Newton iteration to refine the initial conditions and minimize the present-day phase-space mismatch. Table~\ref{tab:LMC_orbit_variation} compares the resulting velocity statistics between the original neural network-derived trajectories and the refined solutions.

The table demonstrates that while most summary statistics remain stable (variations $\lesssim 1$ km/s), the mean radial velocity of stars in the southern hemisphere shows enhanced sensitivity to trajectory uncertainties, with variations reaching 5-8 km/s at 60-90 kpc. This finding motivated our recommendation in Section~\ref{sec:limitations} to either down-weight or exclude this particular statistic in regions where trajectory uncertainties dominate.

\renewcommand{\arraystretch}{1.3}
\setlength{\tabcolsep}{4.pt}
\onecolumn
\captionsetup{width=\textwidth}
\begin{longtable}{ccccccc}
\caption{Variations in stellar kinematic summary statistics due to LMC trajectory reconstruction uncertainties. For five MW-LMC parameter combinations showing the largest phase-space deviations, we compare results from neural network-derived trajectories (before slash) with Gauss-Newton refined trajectories (after slash). The table shows: (1) initial conditions at $t=-2$ Gyr when the LMC enters the MW virial radius, (2) final conditions at present day, and (3) absolute changes in mean velocities and velocity dispersions across three radial bins. Most statistics vary by $\lesssim 1$ km/s, except $\langle v_{r,b<0^{\circ}} \rangle$ at 60-90 kpc, which can change by up to 8 km/s, highlighting its sensitivity to trajectory uncertainties.}\label{tab:LMC_orbit_variation} \\
    \hline
    \multicolumn{7}{c}{$M_{\mathrm{MW}}=0.86\times10^{12}{\rm M}_{\odot}$, $M_{\mathrm{LMC}}=1.88\times10^{11}{\rm M}_{\odot}$, $c=10.11$ and $q=1.14$} \\
    \hline
    & $\mathrm{X}$ (kpc) & $\mathrm{Y}$ (kpc) & $\mathrm{Z}$ (kpc) & $\mathrm{V_{x}}$ (km/s) & $\mathrm{V_{y}}$ (km/s) & $\mathrm{V_{z}}$ (km/s)\\
    \hline
    Initial conditions & 24.1 / 27.27  & 279.81 / 268.83 & 87.43 / 67.57 & 1.2 / -1.44 & -69.17 / -61.80  & -80.27 / -77.06 \\
    \hline
    Final conditions & -1.09 / -2.50 & -37.17 / -38.74 & -29.11 / -27.32 & -59.13 / -48.19 & -248.36 / -226.94 & 183.48 / 203.23 \\
    \hline
    & $\mathrm{\Delta\sigma_{v_r}}$ (km/s) & $\mathrm{\Delta\sigma_{v_b}}$ (km/s) & $\mathrm{\Delta\sigma_{v_l}}$ (km/s) & $\mathrm{\Delta\langle v_{r,b>0^{\circ}} \rangle}$ (km/s) & $\mathrm{\Delta\langle v_{r,b<0^{\circ}} \rangle}$ (km/s) & $\mathrm{\Delta\langle v_{b} \rangle}$ (km/s)\\
    \hline
    30-60 kpc & 0.3 & 0.1 & 0.0 & 0.3 & 1.0 & 0.1\\
    \hline
    60-90 kpc & 0.0 & 0.0 & 0.3 & 0.3 & 3.1 & 0.5\\
    \hline
    90-120 kpc & 0.3 & 0.2 & 0.2 & 1.1 & 1.0 & 1.1 \\
    \hline
    \multicolumn{7}{c}{$M_{\mathrm{MW}}=0.80\times10^{12}{\rm M}_{\odot}$, $M_{\mathrm{LMC}}=1.15\times10^{11}{\rm M}_{\odot}$, $c=10.41$ and $q=1.11$} \\
    \hline
    & $\mathrm{X}$ (kpc) & $\mathrm{Y}$ (kpc) & $\mathrm{Z}$ (kpc) & $\mathrm{V_{x}}$ (km/s) & $\mathrm{V_{y}}$ (km/s) & $\mathrm{V_{z}}$ (km/s)\\
    \hline
    Initial conditions & 26.42 / 24.84  & 261.43 / 259.41 & 27.33 / 23.80 & 2.06 / 3.29 & -63.73 / -65.72  & -62.87 / -66.27 \\
    \hline
    Final conditions & -0.96 / -0.57 & -37.19 / -41.49 & -24.9 / -27.85 & -55.34 / -53.95 & -203.67 / -193.72 & 218.99 / 210.53 \\
    \hline
    & $\mathrm{\Delta\sigma_{v_r}}$ (km/s) & $\mathrm{\Delta\sigma_{v_b}}$ (km/s) & $\mathrm{\Delta\sigma_{v_l}}$ (km/s) & $\mathrm{\Delta\langle v_{r,b>0^{\circ}} \rangle}$ (km/s) & $\mathrm{\Delta\langle v_{r,b<0^{\circ}} \rangle}$ (km/s) & $\mathrm{\Delta\langle v_{b} \rangle}$ (km/s)\\
    \hline
    30-60 kpc & 0.8 & 0.2 & 0.4 & 0.6 & 2.7 & 1.1\\
    \hline
    60-90 kpc & 0.2 & 0.1 & 0.0 & 1.1 & 4.7 & 1.4\\
    \hline
    90-120 kpc & 0.4 & 0.2 & 0.1 & 1.4 & 1.1 & 2.0 \\
    \hline
    \multicolumn{7}{c}{$M_{\mathrm{MW}}=1.40\times10^{12}{\rm M}_{\odot}$, $M_{\mathrm{LMC}}=1.79\times10^{11}{\rm M}_{\odot}$, $c=13.02$ and $q=0.64$} \\
    \hline
    & $\mathrm{X}$ (kpc) & $\mathrm{Y}$ (kpc) & $\mathrm{Z}$ (kpc) & $\mathrm{V_{x}}$ (km/s) & $\mathrm{V_{y}}$ (km/s) & $\mathrm{V_{z}}$ (km/s)\\
    \hline
    Initial conditions & 24.92 / 23.71  & 269.91 / 274.52 & 86.01 / 109.12 & 5.61 / 6.71 & -37.15 / -42.06  & -91.09 / -93.40 \\
    \hline
    Final conditions & -0.95 / -1.13 & -42.41 / -44.66 & -30.01 / -29.23 & -56.34 / -62.70 & -248.4 / -259.98 & 255.81 / 241.50 \\
    \hline
    & $\mathrm{\Delta\sigma_{v_r}}$ (km/s) & $\mathrm{\Delta\sigma_{v_b}}$ (km/s) & $\mathrm{\Delta\sigma_{v_l}}$ (km/s) & $\mathrm{\Delta\langle v_{r,b>0^{\circ}} \rangle}$ (km/s) & $\mathrm{\Delta\langle v_{r,b<0^{\circ}} \rangle}$ (km/s) & $\mathrm{\Delta\langle v_{b} \rangle}$ (km/s)\\
    \hline
    30-60 kpc & 0.2 & 0.2 & 0.1 & 0.3 & 0.3 & 0.2\\
    \hline
    60-90 kpc & 0.1 & 0.6 & 0.1 & 0.1 & 1.0 & 0.6\\
    \hline
    90-120 kpc & 0.5 & 0.5 & 0.3 & 0.8 & 0.9 & 0.8 \\
    \hline
    \multicolumn{7}{c}{$M_{\mathrm{MW}}=0.61\times10^{12}{\rm M}_{\odot}$, $M_{\mathrm{LMC}}=0.74\times10^{11}{\rm M}_{\odot}$, $c=7.90$ and $q=1.44$} \\
    \hline
    & $\mathrm{X}$ (kpc) & $\mathrm{Y}$ (kpc) & $\mathrm{Z}$ (kpc) & $\mathrm{V_{x}}$ (km/s) & $\mathrm{V_{y}}$ (km/s) & $\mathrm{V_{z}}$ (km/s)\\
    \hline
    Initial conditions & 38.45 / 33.32  & 271.0 / 314.06 & -20.33 / 8.33 & -4.82 / -10.25 & -97.21 / -120.63  & -39.58 / -43.10 \\
    \hline
    Final conditions & -0.76 / -2.99 & -40.73 / -40.42 & -30.75 / -26.51 & -60.06 / -37.28 & -182.88 / -220.09 & 160.19 / 150.50 \\
    \hline
    & $\mathrm{\Delta\sigma_{v_r}}$ (km/s) & $\mathrm{\Delta\sigma_{v_b}}$ (km/s) & $\mathrm{\Delta\sigma_{v_l}}$ (km/s) & $\mathrm{\Delta\langle v_{r,b>0^{\circ}} \rangle}$ (km/s) & $\mathrm{\Delta\langle v_{r,b<0^{\circ}} \rangle}$ (km/s) & $\mathrm{\Delta\langle v_{b} \rangle}$ (km/s)\\
    \hline
    30-60 kpc & 1.5 & 0.6 & 0.5 & 0.5 & 3.5 & 1.5\\
    \hline
    60-90 kpc & 0.0 & 0.6 & 0.8 & 0.8 & 7.5 & 1.5\\
    \hline
    90-120 kpc & 0.0 & 0.3 & 0.4 & 0.4 & 1.6 & 2.1\\
    \hline
    \multicolumn{7}{c}{$M_{\mathrm{MW}}=0.82\times10^{12}{\rm M}_{\odot}$, $M_{\mathrm{LMC}}=1.40\times10^{11}{\rm M}_{\odot}$, $c=11.22$ and $q=1.10$} \\
    \hline
    & $\mathrm{X}$ (kpc) & $\mathrm{Y}$ (kpc) & $\mathrm{Z}$ (kpc) & $\mathrm{V_{x}}$ (km/s) & $\mathrm{V_{y}}$ (km/s) & $\mathrm{V_{z}}$ (km/s)\\
    \hline
    Initial conditions & 28.18 / 11.66  & 249.83 / 256.39 & 29.97 / 65.18 & 3.57 / 7.50 & -51.91 / -52.85  & -68.04 / -73.68 \\
    \hline
    Final conditions & -1.26 / 0.66 & -37.65 / -38.08 & -28.04 / -27.60 & -57.49 / -50.44 & -196.65 / -221.66 & 221.09 / 200.98 \\
    \hline
    & $\mathrm{\Delta\sigma_{v_r}}$ (km/s) & $\mathrm{\Delta\sigma_{v_b}}$ (km/s) & $\mathrm{\Delta\sigma_{v_l}}$ (km/s) & $\mathrm{\Delta\langle v_{r,b>0^{\circ}} \rangle}$ (km/s) & $\mathrm{\Delta\langle v_{r,b<0^{\circ}} \rangle}$ (km/s) & $\mathrm{\Delta\langle v_{b} \rangle}$ (km/s)\\
    \hline
    30-60 kpc & 0.5 & 0.3 & 0.3 & 0.4 & 0.9 & 0.3\\
    \hline
    60-90 kpc & 0.2 & 0.0 & 0.3 & 0.4 & 2.8 & 0.7\\
    \hline
    90-120 kpc & 0.2 & 0.8 & 0.2 & 0.2 & 1.7 & 0.7\\
    \hline
\end{longtable}
\twocolumn


\bsp	
\label{lastpage}
\end{document}